%% file: festschrift2013.tex
\documentclass[submission,copyright,creativecommons]{eptcs}
\providecommand{\event}{Festschrift 2013} 
\usepackage{breakurl} 

\usepackage{pgf}
\usepackage{tikz}
\usetikzlibrary{arrows,automata}
\usetikzlibrary{positioning}
\usetikzlibrary{calc}
\usetikzlibrary{decorations.markings}
\usepackage{graphicx}
\usepackage{dsfont,stmaryrd}
\usepackage{color,graphicx}
\usepackage{ifthen}
\usepackage{amssymb, amsmath, amsthm}
\usepackage{alltt}
\usepackage{array}
\usepackage[utf8]{inputenc}  
\usepackage[T1]{fontenc}
\usepackage[english]{babel}
\usepackage{xspace}
\usepackage{url}
\usepackage[normalem]{ulem} 
\usepackage{hevea}
\usepackage{varwidth}
\graphicspath{{Figures/}}

\usepackage{caption}

\DeclareCaptionFormat{myformat}{#1#2#3\vspace{-.8em}\hrulefill}
\usepackage{float}
\floatstyle{plaintop}
\restylefloat{figure}

\usepackage{listings}
\lstdefinelanguage{pantagruel} {morekeywords={interface, attribute,
    event, action, rules, when, from, value, changed, with, and, or,
    all, groupby, trigger, on, end}, sensitive=true,
  morecomment=[l]{//}, morecomment=[s]{/*}{*/}, }

\let\oldfigure\figure
\let\oldendfigure\endfigure
\def\figure{\begingroup \oldfigure}
\def\endfigure{\hrulefill\oldendfigure \endgroup}

\input{grammar}
\input{semantics}
\input{denotational}

\def\titlerunning{Denotational semantics of a Domain-Specific Language}
\def\authorrunning{J. Mercadal, Z. Drey, \& C. Consel}

\newcommand{\ie}{\emph{i.e.,}\xspace}
\newcommand{\eg}{\emph{e.g.,}\xspace}

\newcommand{\myparagraph}[1]{\vspace{0.3\baselineskip}\noindent\textit{#1.}}

\tikzset{
    state/.style={
           rectangle,
           rounded corners,
           draw=black, thin,
           minimum height=2em,
           inner sep=1pt,
           text centered,
           },
    comment/.style={rectangle, 
             inner sep= 5pt, 
             text width=4cm, 
             node distance=0.25cm, font=\sffamily},
    invisible/.style={rectangle, fill=white, draw=white, text=black}
}

\newboolean{showcomments}
\setboolean{showcomments}{false}
\ifthenelse{\boolean{showcomments}}
{ \newcommand{\mynote}[2]{
    \fbox{\bfseries\sffamily\scriptsize#1}
    {\small$\blacktriangleright$\textsf{\emph{#2}}$\blacktriangleleft$}}}
{ \newcommand{\mynote}[2]{}}

\title{Denotational Semantics of \\ A User-Oriented, Domain-Specific
  Language}
\author{
  Julien Mercadal 
  \institute{ASCOLA group\\
  INRIA \& Ecole des Mines\\
  Nantes (France)}\\
  \email{julien.mercadal@mines-nantes.fr}
 \and
  Zo\'e Drey
  \institute{STIC-IDM\\
  Lab-STICC \& ENSTA Bretagne\\
  Brest (France)}\\
  \email{zoe.drey@ensta-bretagne.fr}
  \and
  Charles Consel  
  \institute{PHOENIX group\\
  INRIA \& University of Bordeaux\\
Bordeaux (France)}\\
  \email{charles.consel@inria.fr}
}

\begin{document}
\maketitle

\begin{abstract}
  This paper presents the formal definition of a domain-specific
  language, named Pantagruel, following the methodology proposed by
  David Schmidt for language development. This language is dedicated
  to programming applications that orchestrate networked entities. It
  targets developers that are professionals in such domains as
  building management and assisted living, and want to leverage
  networked entities to support daily tasks.

  Pantagruel has a number of features that address the requirements of
  the domain of entity orchestration. Furthermore, Pantagruel provides
  high-level constructs that make it accessible to developers that do
  not necessarily have programming skills. It has been used to develop
  a number of applications by non-programmers.

  We show how the user-oriented programming concepts of Pantagruel are
  expressed in the denotational semantics of Pantagruel. This formal
  definition has been used to derive an interpreter for Pantagruel and
  to provide a basis to reason about Pantagruel programs.
\end{abstract}
\input{intro}
\input{domain}
\input{tour}
\input{denstyle}
\input{speclayer} 
\input{dynsem}
\input{conclu}

\bibliographystyle{eptcs} 
\bibliography{festschrift2013}   

%
\end{document}

%% file: grammar.tex
\ifhevea
\def\NT#1{{\maroon #1\/}}
\def\KW#1{{\blue #1}}
\else
\def\NT#1{{\rm #1\/}}
\def\KW#1{{\bf #1}}
\fi

\ifhevea
\def\gramor{{\black $|$}}   
\def\grameq{{\black \quad::=\quad}}
\def\gramin{{\black \quad$\in$\quad}}
\else
\def\gramor{$\vert$}
\def\grameq{\,\,\,::=\,\,\,\,\,}
\def\gramin{\,\,\,$\in$\,\,\,\,\,}
\fi

\ifhevea

\else

\fi

\def\OR{\gramor\ }

\newenvironment{grammar}{\begin{tabular}{l@{}c@{}l}}{\end{tabular}}
\def\RULE#1\CASE#2{\NT{#1} & \grameq & \KW{#2} \\}
\def\CASE#1{& \gramor & \KW{#1} \\}

\newenvironment{syndom}{\begin{tabular}{l@{}c@{}l}}{\end{tabular}\vspace{0.3cm}\\}
\newcommand{\DOMAIN}[2]{\NT{#1} & \gramin & \NT{#2} \\}

\def\lb{\char123}
\def\rb{\char125}

%% file: semantics.tex
\newcommand{\mita}[1]{\mbox{\it{{#1}}}}

\ifhevea

\else

\fi

\newcommand{\inductrule}{}


%% file: denotational.tex
\usepackage{scalefnt}
\usepackage{setspace} 
\usepackage{listings}
\usepackage{multicol}
\usepackage{marvosym} 
\usepackage{rotating} 
\usepackage{comment}
\usepackage{float}
\usepackage{titlesec}
\usepackage{xifthen}
\usepackage{paralist} 
\newlength\myindent
\newcounter{domcounter}
\newcounter{domtitle}[domcounter]
\newfont{\DomainTitlefnt}{ptmb8t at 11pt}

\newcommand{\kw}[1]{\textbf{#1}}

\newenvironment{SemanticAlgebra}{\begin{minipage}{\textwidth}}
  {\end{minipage}\vspace{.5em}\vspace{.5em}}
\newcounter{DomainTitle}
\newcommand{\DomainTitle}[1]{\vspace{1em}\refstepcounter{domtitle}{%
\noindent\normalsize \MakeUppercase{\roman{domtitle}}.~#1}\par}

\newcommand{\DomainDef}[3]{\par Domain \mita{#1 $\in$ #2 $=$ #3}} 
\newcommand{\DomainDefNoOp}[3]{\par Domain \mita{#1 $\in$ #2 $=$ #3}\par} 
\newcommand{\DomainDefInline}[3]{Domain \mita{#1 $\in$ #2 $=$ #3}} 

\newcommand{\OPS}{\par Operations\par}

\newcommand{\OperationDef}[3]{\par\textit{#1: #2 \\\mita{~~#1}= #3}}
\newcommand{\OperationInfixDef}[5]{\textit{#1: #4 \\#2\mita{~~#1} #3} = #5 \par}
\newcommand{\OperationDefMath}[3]{\textit{#1: #2} \\\mita{~~#1= }\IndentedFormula{#3}\par}
\newcommand{\DN}[1]{\textnormal{{\it #1}}}
\newcommand{\DO}[1]{
  \ifthenelse{\equal{#1}{equals}}{~\DN{#1}~}{
  \ifthenelse{\equal{#1}{join}}{~\DN{#1}~}{%
  \ifthenelse{\equal{#1}{and}}{~\DN{#1}~}{%
  \ifthenelse{\equal{#1}{neq}}{~\DN{#1}~}{\DN{#1}}}}}}

\newcommand{\SetOfMath}[3][]{\{#2 ~|~ #3\}\ifthenelse{\isempty{#1}}{}{_{#1}}}
\newcommand{\SeqOfMath}[3][]{[~#2 ~|~ #3 ~]\ifthenelse{\isempty{#1}}{}{_{#1}}}
\newcommand{\Set}[2][]{\{ #2 \ifthenelse{\isempty{#1}}{}{_{#1}}\}}
\newcommand{\Seq}[2][]{[ #2 \ifthenelse{\isempty{#1}}{}{_{#1}}]}

\newcommand{\LetMath}[2]{\textnormal{\preg{let} } #1 = #2 \ }
\newcommand{\InS}[1]{\textnormal{\preg{in} } #1}
\newcommand{\In}[1]{\\%
{\setlength\arraycolsep{\myindent}\begin{array}[t]{l}\InS{#1}\end{array}}}
\newcommand{\IndentedFormula}[1]{{\setlength\arraycolsep{2.5em}\begin{array}[t]{@{}l}#1\end{array}}}

\newcommand{\IfThen}[2]{#1 \mto #2}

\newcommand{\Where}[2]{\par\hspace{1em}\xwhere~ \mita{#1} = \mita{#2}}
\newcommand{\WhereIndentedFormula}[2]{\par \xtab\xwhere\IndentedFormula{#1 = #2}}

\newenvironment{ValuationFunctions}{
	\smallskip\par\begin{array}[b]{l}}{\end{array}\vspace{0.5em}\par}

\newcommand{\infuncomp}[1]{\ensuremath{\llbracket}\textnormal{#1}\ensuremath{\rrbracket}}
\newcommand{\infuncompi}[2]{\ensuremath{\llbracket}\textnormal{#1}\ensuremath{_{#2}\rrbracket}}
\newcommand{\FC}[1]{{\bf #1}\infuncomp{#1}  } 
\newcommand{\FCI}[2]{{\bf #1}\infuncompi{#1}{#2}} 
\newcommand{\FCOMP}[2]{{\bf #1}\infuncomp{#2}  } 
\newcommand{\FID}[1]{\infuncomp{#1}}
\newcommand{\FIDp}[1]{\infuncomp{\pre{#1}}}
\newcommand{\FIDI}[2]{\infuncompi{#1}{#2}}
\newcommand{\FunDecl}[3] {{\bf{#1}}: \mita{\textnormal{#2} {\!$\to\!$} #3} }
\newcommand{\FunDef}[4][]{\vspace{1.4pt}\hspace{0.5em}{\bf{#2}} \infuncomp{#3} = #4 #1}
\newcommand{\seq}[1]{#1$_1$; #1$_2$}
\newcommand{\FunSeq}[3][]{\vspace{1.4pt}\hspace{0.5em}{\bf{#2}}\infuncomp{\seq{#2}}={#3} #1}
\newcommand{\IN}[2]{\textnormal{in\DN{#1}}(#2)}
\newcommand{\IS}[2]{\textnormal{is\DN{#1}}(#2)}
\newcommand{\param}[1]{\ensuremath{\lambda#1}.}

\newcommand{\dasub}[1]{\!\!\downarrow\!\!_{#1}}
\newcommand\otherwise{\,[]\,}

\newcommand\pre[1]{\texttt{#1}}

\newcommand\preg[1]{\selectfont\sffamily #1}
\newcommand{\PowerSet}[1]{\ensuremath{\mathcal{P}(}\mita{#1}\ensuremath{)}}
\newcommand\mto{\ensuremath{\to}\ }\newcommand\mtimes{\ensuremath{\times}\ }

\newcommand\mtup[1]{\ensuremath{\langle #1 \rangle}}
\newcommand{\xlet}{\textnormal{\;\preg{let}\ }}
\newcommand{\xcases}{\textnormal{\preg{cases}\ }}
\newcommand{\xchoice}[2]{#1 \mto #2}
\newcommand{\xend}{\textnormal{\;\preg{end}\ }}
\newcommand{\xdomain}{\textnormal{\;\preg{domain}}}
\newcommand{\xelse}{\ensuremath{\otherwise}}
\newcommand{\xof}{\textnormal{\;\preg{of}\ }}
\newcommand{\xstr}[1]{\lceil\textnormal{#1}\rceil}
\newcommand{\xin}{\textnormal{\;\preg{in}\ }}
\newcommand{\xwhere}{\textnormal{\;\preg{where}\ }}
\newcommand{\xand}{\textnormal{\;\preg{and}\ }}
\newcommand{\xtab}{\hspace{2em}}
\newcommand{\xdots}[1]{\cdot\cdot\cdot#1\cdot\cdot\cdot}
\newcommand{\xnote}[1]{\textnormal{\small #1}}

\newcommand{\infunmap}[3]{\boldsymbol{ [} #1 \mapsto #2 \boldsymbol{ ]} #3 }
\newcommand{\callupdate}[3]{\infunmap{#1}{#2}{#3}}

\newcommand\NN{\mathds N}
\newcommand\BB{\mathds B}

\newcommand\NOTE[1]{{\small\ \ \textnormal{(#1)}}}
\newcommand\SEMNOTE[1]{{\small\ \ \textnormal{#1}}}



%% file: intro.tex
\section{Introduction}
The realisation of domain-specific languages (DSLs) contrasts with
that of general-purpose programming languages (GPLs) where generality, 
expressivity and power are expected. Instead, a DSL revolves around a
narrow, specific application domain and its users, providing
high-level abstractions and constructs tailored towards this domain
and the needs of users~\cite{conseldsl, martinfowler,
  little-languages, annotated}. Thus, much time and effort must be
devoted to the domain analysis and the language design. Methodologies
and tools have been proposed to develop DSLs, covering the complete
development life-cycle, from design~\cite{conselmarlet, dsel} to
implementation~\cite{whenandhow, notablepatterns, tratt08, wangdsl},
including formalisation~\cite{liang96moddensem, davidschmidt}.

If such methodologies are well-known in the programming-language
community, they hardly have an echo in the end-user community yet. As
an illustration, in their state-of-the-art paper about end-user
software engineering, Ko \emph{et. al}~\cite{endusersoftware} address
the challenge of enabling the end user to produce reliable programs
that actually achieve his requirements, by supporting the programming
task with visualization and simulation tools.
However, they do not mention the need for a formal definition of an
end-user language so that the supporting tools be provably consistent
with the language definition.
A first step towards addressing this issue is
to bridge the gap between the domain analysis, taking the requirements
into account, and the design of a language, providing the constructs
to achieve these requirements.

In this paper, we present the formal definition of a domain-specific
language, named Pantagruel~\cite{jvlc12, dsl09}, following the
methodology proposed by David Schmidt for language
development~\cite{davidschmidt}.
This language is dedicated to programming applications that
orchestrate networked entities. It targets end users that are
professionals, in areas such as building management and assisted
living, and leverages networked entities to support daily tasks. For
example, security guards are assisted by an anti-intrusion system that
coordinates motion detectors and surveillance cameras to signal the
presence of an intruder on monitoring screens. For another example, a
range of assistive devices, like time trackers, task prompters, and
motion sensors, are available to support caregivers in assisting
people with disabilities in their daily life. Pantagruel offers a
number of features that address the requirements of the domain of
networked entity orchestration, while making it accessible to users
that do not necessarily have programming skills. Specifically,
Pantagruel provides two language layers, one for specifying interfaces
from which entities are manipulated, and another for orchestrating
entities. The orchestration layer offers high-level operators taking
advantage of the information defined in the specification layer to
discover, select concrete entities, and interact with them via their
interfaces.

This paper gives a definition of Pantagruel in the form of a
denotational semantics, exhibiting the domain-specific features of
this language. In particular, this formal definition makes the key
concepts of the orchestration of networked entities (\ie entity
discovery and interactions) explicit. It also brings out some
traditional programming language concepts (\eg the notion of memory
and loop, or variable assignment), which have been abstracted away
from the developer.

\subsection*{Outline}
The rest of this paper is organized as follows.
Section~\ref{sec:domain} presents the specificities and the key
concepts of the domain of the orchestration of networked
entities. Section~\ref{sec:tour} gives a tour of our DSL, Pantagruel,
illustrated by a working example of a building-automation
application. Section~\ref{sec:denstyle} motivates the use of the
denotational style for our language.  Sections~\ref{sec:speclayer}
and~\ref{sec:dynsem} describe the syntax and the denotational
semantics of the specification layer and of the orchestration layer of
Pantagruel, respectively. Conclusion and future work are given in
Section~\ref{sec:conclu}.












%% file: domain.tex
\section{Orchestration of networked entities}
\label{sec:domain}
A wide variety of networked entities, both hardware and software, are
populating smart spaces that become prevalent in a growing number of
areas, including supply chain management, building automation,
healthcare, and assisted living. These entities have {\em sensing
  capabilities}, enabling to collect data (\eg ambient temperature, or
user presence), and/or {\em actuating capabilities}, enabling to
perform actions (\eg turn on/off an entity, or display a
message). Additionally, they are characterized by {\em attributes}
(\eg a location or a status). Applications need to be developed to
exploit the capabilities provided by these entities. Such applications
determine the actions to be triggered according to sensed data and
values of entity attributes. We refer to such applications as {\em
  orchestration applications}. The development of orchestration
applications is challenging, requiring to cope with heterogeneity and
dynamicity of networked entities.

\myparagraph{Heterogeneity} The entities of a smart space are
off-the-shelf software components and devices that use a variety of
communication protocols and rely on intricate distributed systems
technologies. The heterogeneity of these entities and the intricacies
of underlying distributed technologies tend to percolate in the code
of the orchestration logic, cluttering it with low-level details.
To facilitate the development of the orchestration logic, it is
necessary to raise the level of abstraction at which entities are
manipulated. To do so, an abstraction layer between the entity
implementation and the orchestration logic must be defined. We refer
to such a layer as a {\em specification layer}, enabling to describe
entities according to their sensing and actuating capabilities and
their attributes. This layer is coupled with the {\em orchestration
  layer} 
%
enabling to write an orchestration
application with respect to the entity descriptions from the
specification layer.  Specifically, an orchestration application
consists of a set of rules that specify which actions on entities need
to be performed (the right part of rules) when some data are sensed by
entities (the left part of rules).

\myparagraph{Dynamicity} An application may interact with a changing
set of entities because they may become (1) available after the
application is deployed, (2) unavailable due to malfunction (\eg power
loss), or (3) unreachable due to a network failure. The code of the
orchestration logic should manage these variations at a high level,
abstracting over current entities of a given smart space.  To do so, we
define an abstraction for interacting with a set of entities sharing
some capabilities and attributes, regardless of the current entities of
the smart space. We refer to such an abstraction as an {\em
  interface}. An interface groups together a set of sensing and
actuating capabilities and attributes to which entity implementations
must conform. Interfaces may be organized hierarchically, enabling
them to inherit capabilities and attributes.

We have built a domain-specific language, named
Pantagruel~\cite{jvlc12, dsl09}, for developing orchestration
applications using the methodology for language development proposed
by Schmidt~\cite{davidschmidt}.  Pantagruel consists of two language
layers: a specification layer for describing the entity interfaces,
and an orchestration layer for describing the orchestration logic. The
syntax and denotational semantics of both layers are presented in the
following sections. Notably, Pantagruel relies on a pervasive
computing platform, named DiaSuite~\cite{diaspectse}, for dealing with
the variety of underlying, low-level technologies.






%% file: tour.tex
\section{Tour of Pantagruel}
\label{sec:tour}
The Pantagruel language is introduced using a working example: a
building-automation application. This application is responsible for
switching on/off lights of a room when a motion is detected in a
room. It also helps regulating the temperature of rooms using
fans. Although rather simplistic, this working example is sufficient
to illustrate the salient features of the Pantagruel language.  More
elaborate examples can be found elsewhere~\cite{phddrey}.

A key feature of Pantagruel is the tight coupling of the
specification layer with the orchestration layer. It allows users to
easily and safely develop their own orchestration
applications. Figure~\ref{fig:specification-layer} gives a
specification for our working example. This specification consists of
four entity interfaces and eight entity instances. For example, the
\texttt{Motion\-Detector} interface declares a sensing capability,
named \texttt{detected}, enabling to signal the presence or the
absence of a motion. The \texttt{Light} interface declares an
actuating capability, named \texttt{switch}, enabling to switch on/off
the entity. Both interfaces have a \texttt{room} attribute, indicating
the location of the entity in the building. For simplicity, the types
of all the elements in this specification are either integer or
boolean. Next, entity instances are defined, according to the
declarations of interfaces. For example, \texttt{l10} is an entity
whose implementation conforms to the \texttt{Light} interface. As a
result, this entity can be switched on/off. Moreover, its
\texttt{room} attribute is initialized to the value \texttt{101}.
\begin{figure}[htbp]
\begin{minipage}[t]{.48\textwidth}
\texttt{\underline{Interfaces}} 
\begin{lstlisting}[language=pantagruel,
      breakatwhitespace=true]
interface MotionDetector {
      attribute room : Integer
      event detected : Boolean  }
interface Light {
      attribute room : Integer
      action switch( Boolean ) }
interface Fan {
      attribute room : Integer
      action setSpeed( Integer ) }
interface TemperatureSensor {
      event temperature : Integer }
\end{lstlisting}
\end{minipage}
\vline\vline\hspace{2em}
\begin{minipage}[t]{.40\textwidth}
\texttt{\underline{Entities}} 
\begin{lstlisting}[language=pantagruel,
      breakatwhitespace=true]
m10:MotionDetector { room : 101 }
m20:MotionDetector { room : 201 }

l10:Light { room : 101 }
l11:Light { room : 101 }
l20:Light { room : 201 }

fan10:Fan { room : 101 }
fan20:Fan { room : 201 }

thermo:TemperatureSensor{}
\end{lstlisting}
\end{minipage}
\caption{A specification for a building equipped with motion
  detectors, lights, fans, and temperature sensors}
\label{fig:specification-layer}
\end{figure}

Given a specification, an orchestration logic is then defined to
orchestrate the entities. In Pantagruel, this orchestration logic
consists of a set of rules, combining the sensing and actuating
capabilities of entities. Figure~\ref{fig:orchestration-layer} gives
three orchestration rules for our working example. Rule~(1) switches
the lights on (referred to as \texttt{l}) in the rooms where a motion
has been detected (instance of motion detector referred to as
\texttt{m}). Inversely, Rule~(2) switches the lights off in the rooms
where no motion has been detected. This rule uses the operators
\textbf{all} and \textbf{groupby} to aggregate and group together some
data collected by entities. These domain-specific operators are not
detailed in this paper. Finally, Rule~(3) adjusts the
speed of a fan when two conditions are satisfied : (1) a light in the
room where the fan is located has been switched on, and (2) the
temperature sensed by the \texttt{thermo} entity is equal to
\texttt{30}.  This rule illustrates the ability of an orchestration
logic to react to the execution of an action (here, a light is
switched on). To do so, each action in Pantagruel produces an {\em
  implicit event} when it is triggered. This event has the same name
as the action and its type consists of the types of the action
parameters. These three rules show how the language constructs enable
to interact either with a particular entity (\eg the \texttt{thermo}
entity in Rule (3)), or with a set of entities via their
interfaces. Moreover, the dependencies between the entities within an
orchestration rule are made explicit by the \textbf{with} clause.
\begin{figure}[htbp]
\begin{lstlisting}[language=pantagruel,
      breakatwhitespace=true]
rules
(1) when 
       event detected from m:MotionDetector value = true 
     trigger 
       action switch(true) on l:Light with room = m.room
    end

(2) when 
       all event detected from m:MotionDetector value = false groupby room 
     trigger 
       action switch(false) on l:Light with room = m.room
    end

(3) when 
       event switch from l:Light value = true
       and  event temperature from thermo value = 30 
    trigger 
       action setSpeed(10) on f:Fan with room = l.room
    end   
end
\end{lstlisting}
\vspace{-1.4em}
\caption{An orchestration logic for a building-automation application}
\label{fig:orchestration-layer}
\end{figure}

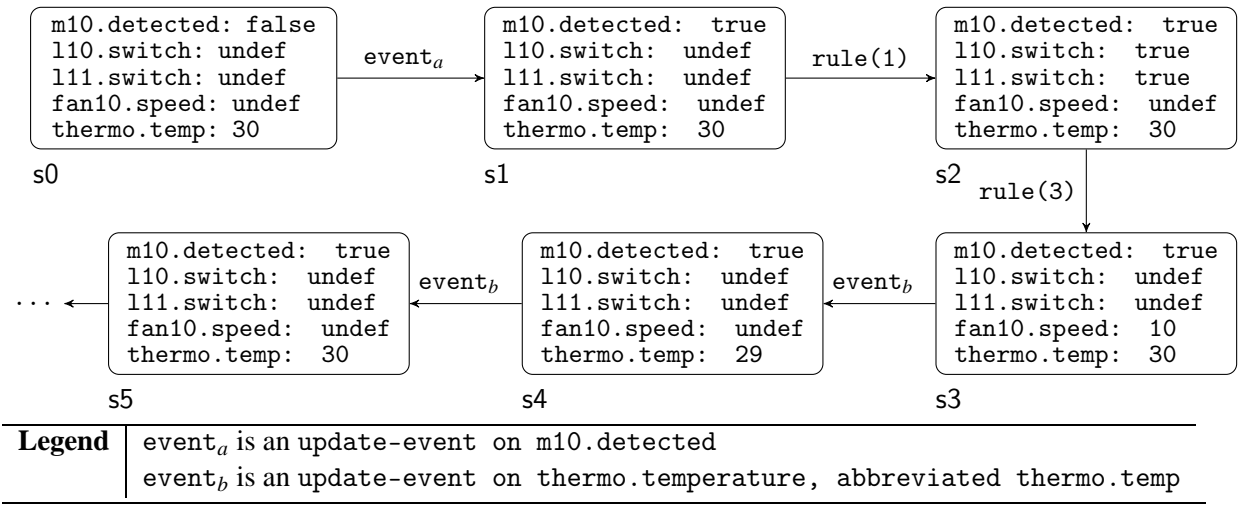
\begin{figure}[htbp]
\input{statechart.tex}
\caption{An example of an execution trace of a Pantagruel program}
\label{fig:statemachine}
\end{figure}

To illustrate the behavior of a Pantagruel program,
Figure~\ref{fig:statemachine} shows an excerpt of an execution trace
for the orchestration rules given in
Figure~\ref{fig:orchestration-layer}.  This trace illustrates the
behavior of our working example when a person enters the room
\texttt{101} which was empty until then.
It is represented as a sequence of states, where
each state consists of both the values of data collected by entities
and the implicit events. The transitions between states correspond
either to an update of collected data, or to a rule
execution. Consider Rule~(1) given in
Figure~\ref{fig:orchestration-layer}. When a person is detected in the
room \texttt{101}, the \texttt{m10} entity updates its
\texttt{detected} data, leading to the state \texttt{s1}. This
situation then satisfies the execution conditions of Rule~(1) for
\texttt{m = m10}, triggering the \texttt{switch} action on the
\texttt{l10} and \texttt{l11} lights. This leads to the state
\texttt{s2}, which contains the updated values for the implicit event
\texttt{switch} of these lights. At this state, the execution
conditions of Rule~(3) are verified for \texttt{thermo} (\ie
\texttt{temperature} value is 30), setting the speed of the
\texttt{fan10} entity. Note that the value of an implicit event is
initially undefined (\texttt{undef}), and after each update caused by
an action execution, it becomes again undefined at the next state
change. From the state \texttt{s3}, the temperature sensed by the
\texttt{thermo} entity is updated twice.  Note that Rule (1) is
triggered only once in the trace, \ie as long as the \texttt{detected}
attribute of the motion detector remains true.  The execution
continues indefinitely, either waiting for updates of data collected
by entities, or the appearance of new entities.
 
%% file: statechart.tex
\begin{tikzpicture}[->,>=stealth']
\tikzstyle{every node}=[font=\small\ttfamily]
 \node[state,    	
  text width=4cm, 	
  yshift=1cm, 		
  anchor=center] (S0) 	
 {\begin{tabular}{l} 	
m10.detected: false\\[-.2em]
l10.switch: undef\\[-.2em]
l11.switch: undef\\[-.2em]
fan10.speed: undef\\[-.2em]
thermo.temp: 30\\
 \end{tabular}
 };
\node[comment, below=0.01 of S0] (comment-S0) {s0\\};

\node[state,
  right of=S0,
  node distance=6cm,
  anchor=center] (S1) 
 {%
 \begin{tabular}{l}
m10.detected: true\\[-.2em]
l10.switch: undef\\[-.2em]
l11.switch: undef\\[-.2em]
fan10.speed: undef\\[-.2em]
thermo.temp: 30\\
 \end{tabular}
 };
\node[comment, below=0.01 of S1] (comment-S1) {s1\\};
\node[state,
  right of=S1,
  node distance=6cm,
  anchor=center] (S2) 
 {%
 \begin{tabular}{l}
m10.detected: true\\[-.2em]
l10.switch: true\\[-.2em]
l11.switch: true\\[-.2em]
fan10.speed: undef\\[-.2em]
thermo.temp: 30\\
 \end{tabular}
 };
\node[comment, below=0.01 of S2] (comment-S2) {s2\\};

\node[state,
  below of=S2,
  node distance=3cm,
  anchor=center] (S3) 
 {%
 \begin{tabular}{l}
m10.detected: true\\[-.2em]
l10.switch: undef\\[-.2em]
l11.switch: undef\\[-.2em]
fan10.speed: 10\\[-.2em]
thermo.temp: 30\\
 \end{tabular}
 };
\node[comment, below=0.01 of S3] (comment-S3) {s3\\};

\node[state,
  left of=S3,
  node distance=5.5cm,
  anchor=center] (S4) 
 {%
 \begin{tabular}{l}
m10.detected: true\\[-.2em]
l10.switch: undef\\[-.2em]
l11.switch: undef\\[-.2em]
fan10.speed: undef\\[-.2em]
thermo.temp: 29\\
 \end{tabular}
 };
\node[comment, below=0.01 of S4] (comment-S4) {s4\\};

\node[state,
  left of=S4,
  node distance=5.5cm,
  anchor=center] (S5) 
 {%
 \begin{tabular}{l}
m10.detected: true\\[-.2em]
l10.switch: undef\\[-.2em]
l11.switch: undef\\[-.2em]
fan10.speed: undef\\[-.2em]
thermo.temp: 30\\
 \end{tabular}
 };
\node[comment, below=0.01 of S5] (comment-S5) {s5};

\node[invisible, left of=S5, node distance=3cm, anchor=center] (none) {...};
 \path (S0) edge[right]  node[anchor=south,above]{event$_a$\ \ } (S1)
 (S1)     	edge[right] node[anchor=south,above]{rule(1)} (S2)
 (S2)      	edge[]       node[anchor=south,left]{rule(3)}      (S3)
 (S3)      	edge[left]     node[anchor=south,above]{event$_b$\ \ }    (S4)
 (S4)      	edge[left]     node[anchor=south,above]{event$_b$\ \ }    (S5)
 (S5)       edge[left]     node[anchor=south,above]{} (none);

\end{tikzpicture}
\begin{tabular}{l|l}
\hline
\textbf{Legend} &
\texttt{event$_a$} is an \texttt{update-event on m10.detected}\\
& \texttt{event$_b$} is an \texttt{update-event on thermo.temperature,
abbreviated thermo.temp}\\
\end{tabular}

%% file: denstyle.tex
\clearpage
\section{Pantagruel in the denotational style}
\label{sec:denstyle}
Defining the semantics of a language
usually relies on one of the following three methodologies:
operational, axiomatic, or denotational~\cite{schmidtintro}.
Denotational semantics focuses on the mathematical relation between
input data and output data, and makes
explicit the domain definition of these data. In our approach, the
structure and characteristics of the data related to the entities are
central to the dynamic
semantics of our language.  Furthermore, because the denotational
semantics uses a functional style, a definition can easily be mapped
into an implementation. In fact, we have implemented in the OCaml
language an interpreter based on the semantics of Pantagruel.

Additionally, the functional style of the denotational semantics
enables us to generalize the definition of orchestration rules, from
an entity to a class of entities.  Specifically, we leverage
higher-order functions to express the rule-evaluation process as a
pattern that is applied on a set of entities, similar to map-like
functions in higher-order languages. In contrast, an operational
definition would make explicit these generalization steps.  We
illustrate this feature
in Section~\ref{sec:orchestration-syntax} with the valuation functions
of rules, events and actions.


%% file: speclayer.tex
\section{The specification layer}
\label{sec:speclayer}
In this section, we present the specification layer of the Pantagruel
language: its semantic algebras, its abstract syntax and its valuation
functions. Let us first introduce some notations used thereafter.

\subsection{Some notations}
This section presents the notations used to define the specification
layer and the orchestration layer of Pantagruel. We use the following
conventions, inspired by Schmidt's methodology~\cite{davidschmidt}:
~\\

$\begin{array}{l@{\hspace{1em}}l}
    \langle a, b \rangle & \textnormal{a tuple of two elements $a$ and $b$} \\
	\infunmap{x}{v}{\rho} & x\textnormal{ maps to } v \textnormal{ in } \rho\\
	x\dasub i, i\in\NN^* & \textnormal{projection on }i^{th} \textnormal{ element
	of tuple } x \textnormal{, where } i \textnormal{ is a non-null integer }\\
	\llbracket X \rrbracket & \textnormal{denotation of a
          syntactic fragment}\\
    T^\omega = \{ t_i \in T \}_{i \in {\mathbb N}} & T^\omega$ is the set of
infinite sequences of elements of set T
$\\T = [ ~ \tau_0, \tau_1, ... ~ ] & T$ is an infinite sequence$\\
\end{array}$
\vspace{1em}~\\
\noindent The last notation is borrowed from ~\cite{BenvenisteGSS92,
  Brogi97modelingcoordination, Kahn74}.  Following Schmidt's
methodology, we define the semantics of the specification layer of
Pantagruel in three steps : first, the semantic algebras are
specified, representing the semantic domains of the language; second,
the abstract syntax of the language is defined; third, the valuation
functions define the meaning of programs, by mapping the abstract
syntax to the semantic domains.

\subsection{Semantic algebras}

The semantic domains for the specification layer of Pantagruel are
listed in Figure~\ref{fig:speclayer-semantic-algebras}.  They show
that the specification layer is mainly understood in terms of
environments mapping identifiers of the language to some value. The
environments are defined in the usual form of a \textit{Map} domain as
illustrated in Figure~\ref{fig:speclayer-environments}, where X stands
for the type of the value stored by the environment.  For clarity, the
\DN{Errvalue} domain used for ill-typed values has been omitted
from Figure~\ref{fig:speclayer-semantic-algebras}.  Let us 
examine the domains of the specification layer.

\begin{figure}[htbp]
\begin{SemanticAlgebra}
\DomainDef{m}{Map}{Id \mto $($X + Errvalue$)$}
\Where{Errvalue}{Unit}
\OPS
\OperationDef{empty$_m$}{Map}{\param{i}$\IN{Errvalue}{}$}
\OperationDef{access$_m$}{Id \mto Map \mto $($X + Errvalue$)$}{\param{i}\param{m}$m(i)$}\\
\OperationDefMath{update$_m$}{Id \mto X \mto Map \mto Map}{\param{i}\param{x}\param{m}\callupdate{i}{\IN{X}{x}}{m}}
\vspace{-.8em}\end{SemanticAlgebra}
\caption{Generic algebra for the definition of environment domains}
\label{fig:speclayer-environments}
\end{figure}

\paragraph{Interface abstraction.} Interfaces are values from the
\DN{Interface} domain (part~\ref{dom:interfaces} of
Figure~\ref{fig:speclayer-semantic-algebras}). To discriminate between
entities, an interface declares typed attributes stored in the
\DN{Interface-attribute} domain. An interface similarly declares its
sensing capabilities in the \DN{Interface-event} domain.  Finally,
actuating capabilities are defined as method signatures, in the
\DN{Interface-action} domain, which maps a method identifier to a
pair: (1) the type of the method parameter and (2) a procedure,
defined by the \DN{Action-struct} domain, to update a store with the
value of an implicit event (see Section~\ref{sec:speclayer-def}).
For conciseness, methods have only one formal parameter.  Interfaces
are stored in an \textit{Env-interface} environment
(part~\ref{dom:env-interface} of
Figure~\ref{fig:speclayer-semantic-algebras}) mapping an identifier to
an interface. This environment is constant for a given Pantagruel
program.

\begin{figure}[htbp]
\begin{SemanticAlgebra}
\vspace{-.8em}
\DomainTitle{Basic domains : truth values, natural numbers, identifiers, and types (operations omitted)}\label{dom:basic}\vspace{.4em}
\begin{minipage}{.5\textwidth}
\DomainDef{t}{\DN{Type}}{\textnormal{Type-structure}}
\DomainDef{b}{Tr}{$\mathbb{B}$}
\end{minipage}\begin{minipage}{.5\textwidth}
\DomainDef{n}{Nat}{$\mathbb{N}$}
\DomainDef{i}{Id}{\textnormal{Identifier}}
\end{minipage}

\DomainTitle{Values}\label{dom:values}
\DomainDef{x}{Value}{Nat + Tr +  Undefinedvalue,  \xwhere UndefinedValue = Unit}
\par
\DomainTitle{Interfaces}\label{dom:interfaces}
\DomainDef{f}{Interface}{Interface-attribute \mtimes Interface-event \mtimes Interface-action}
\WhereIndentedFormula{
  p \in \DN{Interface-attribute}}{\DN{Id} \mto \DN{Type}\\
  s \in \DN{Interface-event} = \DN{Id} \mto \DN{Type}\\
  a \in \DN{Interface-action} = \DN{Id} \mto (\DN{Type} \mtimes \DN{Action-struct})\\
  u \in \DN{Action-struct} = \DN{Id} \mto \DN{Value} \mto \DN{Store} \mto \DN{Store}
}

\DomainTitle{Environments of interfaces}\label{dom:env-interface}
\DomainDef{e}{Env-interface}{Id \mto Interface}
\vspace{-.8em}
\end{SemanticAlgebra}
\caption{Semantic algebras for the specification layer}
\label{fig:speclayer-semantic-algebras}
\end{figure}

\paragraph{Dynamic context data.} Entities are the elements that hold
the runtime information of a Pantagruel program.  Entities are
explicitly represented in the semantics as named objects from the
\DN{Entity} domain; their implementation conforms to one or more
interfaces. For conciseness, we only define entities referring to a
unique interface. Each entity is identified with its current context
data, \ie its attributes and \emph{events}: they conform to the
attribute declarations and to the sensing capabilities of the entity
interface, respectively. Both context data are defined by the
\DO{Entity-attribute} and the \DO{Entity-event} domains.

The \DN{Store} domain (part~\ref{dom:stores} of
Figure~\ref{fig:speclayer-semantic-algebras-store}) holds the set of
available entities indexed by their name, corresponding to the store
variables. When new entities are discovered at runtime, new variables
corresponding to the name of these entities are introduced in the
store.
%
To cumulate the results produced by each rule and, within a rule, by
each action, the \DN{Store} domain further defines the \mita{join}
operation to join two partial stores which are produced either by two rules
or by two actions. The \mita{join} operation is
associative, assuming stores have disjoint
domains, thus suggesting noninterfering parallelism.
As suggested by
Schmidt~\cite{davidschmidt} (Chapter 5), \mita{join} combines the
effects of the execution of actions without overwriting them. This is
achieved by an appropriate definition of a \mita{combine} operation,
enabling partial stores to be combined; it is left unspecified here.

The values of the data sensed by entities, their attributes, and the
actual parameter of actions, are specified by the \textit{Value}
domain (part~\ref{dom:values} of
Figure~\ref{fig:speclayer-semantic-algebras}). The
\textit{Undefinedvalue} domain allows undefined values to be stored.
For simplicity, only natural numbers and truth values
(part~\ref{dom:basic} of Figure~\ref{fig:speclayer-semantic-algebras})
are considered.
\begin{figure}[htbp]
\begin{SemanticAlgebra}
\vspace{-.8em}
\DomainTitle{Entities}

\DomainDef{o}{Entity}{Id \mtimes Entity-attribute \mtimes Entity-event}
\WhereIndentedFormula{
  q \in \DN{Entity-attribute}}{\DN{Id} \mto \DN{Value}\\
  t \in \DN{Entity-event} = \DN{Id} \mto \DN{Value}}

\DomainTitle{Stores}
\label{dom:stores}
\DomainDef{$\sigma$}{Store}{Id \mto Entity$_\bot$}
\OPS
\OperationDef{newstore}{Id \mto Entity$_\bot$}{\param{i}$\bot$}
\OperationDef{join}{Store \mto Store \mto Store}
{\param{\sigma_1}\param{\sigma_2} %
  $(\param{i} \sigma_1(i) ~combine~ \sigma_2(i))$}
\OperationDef{join$^*$}{\PowerSet{Store} \mto Store}
{\param{S} $
\sigma_1 \DO{join} ... (\sigma_{n-1} \DO{join} \sigma_n )$, \textnormal{\hspace{1em}where} 
$\sigma_i \in S, i < 
\left\vert{S}\right\vert$ \textnormal{and} $S$ \textnormal{ is finite}}

\OperationDefMath{\DO{access-action}}{Id \mto Id \mto Env-interface \mto Store \mto Action-struct}{
\param{i_a}\param{i_o}\param{e}\param{\sigma}~(\DO{access}~i_a~(\DO{access}~(\sigma~i_o)\dasub 1~e)\dasub 3)\dasub 2}
\OperationDefMath{\DO{access-event}}{Id \mto Id \mto Store \mto Value}{%
\param{i_e}\param{i_o}\param{\sigma}((\sigma~i_o)\dasub 3~ i_e)}
\OperationDefMath{\DO{update-event}}{Id \mto Id \mto Value \mto Store \mto Store}{%
\param{i_e}\param{i_o}\param{p}\param{\sigma}
\callupdate{i_o}{\mtup{(\sigma~i_o)\dasub 1, (\sigma~i_o)\dasub 2, \callupdate{i_e}{p}{(\sigma~i_o)\dasub 3}}}{\sigma} }
{\small \textnormal{Note : \DO{access-interface}, \DO{access-attribute} and \DO{update-attribute} are handled similarly}
}\vspace{-.5em}
\end{SemanticAlgebra}
\caption{Semantic algebras for the specification layer (continued)}
\label{fig:speclayer-semantic-algebras-store}
\end{figure}

\subsection{Definition of the specification layer}
\label{sec:speclayer-def}

The abstract syntax and the valuation functions for the specification
layer appear in Figures~\ref{fig:speclayer-abstract-syntax} and
~\ref{fig:speclayer-valuation-functions}. A specification S consists
of a set of interfaces F and an initial set of entities O. Note that,
at runtime, these entities can be removed and new entities can be
deployed.
\begin{figure}[h!tbp]
\begin{minipage}[t]{0.48\textwidth}
\begin{syndom}
  \DOMAIN{S}{Specification}
  \DOMAIN{F}{Interface-declaration}
  \DOMAIN{P}{Attribute}
  \DOMAIN{E}{Sensing-capability}
  \DOMAIN{A}{Actuating-capability}
  \DOMAIN{O}{Entity-declaration}
  \DOMAIN{G}{Assignment}
\end{syndom}
\end{minipage}
\begin{minipage}[b]{0.3\textwidth}
\begin{syndom}
  \DOMAIN{T}{Type-structure}
  \DOMAIN{X}{Value}
\end{syndom}
\end{minipage}
\begin{minipage}[b]{0.2\textwidth}
\begin{syndom}
  \DOMAIN{I}{Identifier}
  \DOMAIN{N}{Numeral}
\end{syndom}
\end{minipage}
\begin{minipage}{0.48\textwidth}
\begin{grammar}
  \RULE{S}
  \CASE{\NT{F};\NT{O}}

  \RULE{F}
  \CASE{\NT{F$_1$};\NT{F$_2$} \OR interface \NT{I} \lb\NT{P} ; \NT{E} ; \NT{A}\rb}
  \RULE{P}
  \CASE{\NT{P$_1$};\NT{P$_2$} \OR attribute \NT{I} : \NT{T}}

  \RULE{E}
  \CASE{\NT{E$_1$};\NT{E$_2$} \OR event \NT{I} : \NT{T}}

  \RULE{A}
  \CASE{\NT{A$_1$};\NT{A$_2$} \OR action \NT{I$_1$}( \NT{T} )}

  \RULE{O}
  \CASE{\NT{O$_1$};\NT{O$_2$} \OR entity \NT{I$_o$} : \NT{I$_f$} \lb
    \NT{G}\rb}

  \RULE{G}
  \CASE{\NT{G$_1$};\NT{G$_2$} \OR \NT{I} = \NT{X}}
\end{grammar}
\end{minipage}
\begin{minipage}{0.5\textwidth}
\begin{grammar}
  \RULE{T}
  \CASE{nat \OR bool}
  \RULE{X}
  \CASE{\NT{N} \OR true \OR false}
\end{grammar}
\end{minipage}
\caption{Abstract syntax for the specification layer}
\label{fig:speclayer-abstract-syntax}
\end{figure}  

The \textbf{S} valuation function produces a pair of two environments
named \DN{Env-interface} and \DN{Store}, corresponding to a mapping
from an identifier to an interface, and a mapping from an identifier
to an entity, respectively. This valuation function produces the
environment of entities of a given Pantagruel program by invoking the
\textbf{F} function to produce an environment of interfaces that is
passed as an argument to the \textbf{O} function. In doing so, the
entities can be checked against the interface that they implement.
The \textbf{P}, \textbf{E}, \textbf{A} and \textbf{O} functions are
composed similarly to the \textbf{F} function when involving a
sequence (\eg F$_1$;F$_2$). For conciseness, we have omitted them.
These three functions are responsible for enriching environments. The
\textbf{A} function defines a procedure of the \DN{Action-struct}
domain. Since an action is invoked on an entity, this entity is
referred to as the $i_o$ parameter. Following the procedure
specification suggested by Schmidt~\cite{davidschmidt} (Chapter 8), it
also takes an $x$ formal parameter, which is assumed to be of type
\textbf{T}.  When it is invoked through an entity action, this
procedure generates an implicit event of the name of the concerned
action. Finally, the \textbf{O} function states that the values of
sensed data are initially undefined, and the \textbf{G} function
initializes the attributes to some value.

\begin{figure}[htbp]
\FunDecl{S}{Specification}{Env-interface \mtimes Store}
\begin{ValuationFunctions}
  \FunDef{S}{F O}{\LetMath{e}{\FC{F} \DO{empty}}\InS{\mtup{e, \FC{O}e~\DO{empty}}}}
\end{ValuationFunctions}

\FunDecl{F}{Interface-declaration}{Env-interface \mto Env-interface}
\begin{ValuationFunctions}
  \FunDef{F}{F$_1$;F$_2$}{\FCI{F}{2} \circ \FCI{F}{1}}\\
  \FunDef{F}{\kw{interface} I \kw{\lb} P ; E ; A \kw{\rb}}{\param{e}\DO{update} \FID{I}
    ~\mtup{\FC{P} \DO{empty}, \FC{E} \DO{empty}, \FC{A} \DO{empty}} ~e}
\end{ValuationFunctions}

\FunDecl{P}{Attribute}{Interface-attribute \mto Interface-attribute}
\begin{ValuationFunctions}
  \FunDef{P}{\kw{attribute} I \kw{:} T}{\param{p}\DO{update} \FID{I} ~\FC{T} ~p}
\end{ValuationFunctions}

\FunDecl{E}{Sensing-capability}{Interface-event \mto Interface-event}
\begin{ValuationFunctions}
  \FunDef{E}{\kw{event} I \kw{:} T}{\param{s}\DO{update} \FID{I} ~\FC{T} ~s}
\end{ValuationFunctions}

\FunDecl{A}{Actuating-capability}{Interface-action
  \mto Interface-action}
\begin{ValuationFunctions}
  \FunDef{A}{\kw{action} I \kw{(} T \kw{)}}
  {\param{a}\DO{update} \FID{I} \mtup{\FC{T}, \param{i_o}\param{x}(\DO{update-event}\FID{I}~i_o~x)}~a}
\end{ValuationFunctions}

\FunDecl{O}{Entity-declaration}{Env-interface \mto Store \mto Store}
\begin{ValuationFunctions}
  \FunDef{O}{\kw{entity} I$_{o}$ : I$_{f}$ \kw{\lb} G \kw{\rb}}%
    { \param{e}\param{\sigma}
      \IndentedFormula{
        \DO{update} \FIDI{I}{o}~\mtup{\FIDI{I}{f},%
          (\FC{G}(\DO{access}\FIDI{I}{f}~e)\dasub 1~\DO{empty}), \DO{empty}}~\sigma}
}
\end{ValuationFunctions}

\FunDecl{G}{Assignment}{Interface-attribute \mto  Entity-attribute \mto Entity-attribute}
\begin{ValuationFunctions}
  \FunDef{G}{G$_1$;G$_2$}{\param{p}(\FCI{G}{2}p) \circ (\FCI{G}{1}p)}\\
  \FunDef{G}{I \kw{=} X}{\param{p}\param{q} 
    \DO{update-attribute} \FID{I}~\FC{X}~q}
\end{ValuationFunctions}

\FC{X}\!:\DN{Value} (operations omitted; X is assumed to be a constant numeral or boolean here)
\caption{Valuation functions for the specification layer}
\label{fig:speclayer-valuation-functions}
\end{figure}

In our definitions, type checking is not performed yet. To check that
the value assigned to an attribute is consistent with the interface of
the concerned entity, the functionality of the \textbf{O} function
must be modified to take in account an erroneous \DN{Entity-attribute}
construction. To do so, we use the same definition as the one proposed
by Schmidt for the \DN{Store} domain (see ~\cite{davidschmidt}, Chapter
7): the \DN{Entity-attribute} domain is tagged to indicate
an attribute initialization failure, and a \DO{check} operation
prevents further assignments in case of an error.
\vspace{.6em}\\
\DomainDefInline{Post-entity-attribute}{p}{OK + Err}
\par\hspace{2em}\xwhere \DN{OK} = \DO{Entity-attribute} \DO{and} \DN{Err} = \DO{Entity-attribute}.
\\Operations\\
\noindent\OperationInfixDef{check}{}{f}{$($Entity-attribute \mto Post-entity-attribute$)$\mto $($Entity-attribute \mto Post-entity-attribute$)$}{%
$\param{p}\xcases (p) \xof \xchoice{\IS{OK}{s}}{(f~s)} \xelse \xchoice{\IS{Err}{s}}{p}
\xend$}

~\\\noindent Assignment is then adjusted to check type consistency by calling the \emph{check} operation : ~\\

\noindent\FunDecl{G}{Assignment}{Interface-attribute \mto Entity-attribute
  \mto Post-entity-attribute}
\begin{ValuationFunctions}
  \FunDef{G}{G$_1$;G$_2$}{\param{p} ~(\DO{check}~(\FCI{G}{2}p)) \circ (\FCI{G}{1} p)}\\
  \FunDef{G}{I \kw{=} X}{\param{p}\param{q} 
    \xcases (\DO{access}~\FID{I}~p) \xof\\\hspace{4em} \IndentedFormula{
        \xchoice{\IS{Nat-type}{}}{\xcases x \xof\\\hspace{2em}
          \xchoice{\IS{Nat}{n}}{\IN{OK}{(\DO{update} \FID{I}~(\FC{X})~q)}}
          \xelse
          \xchoice{\IS{Tr}{b}}{\IN{Err}{q}} \xend\\
       }
        \xchoice{\IS{Tr-type}{}}{ ...~\textnormal{similar processing as above}~...}
      \xend}}
\end{ValuationFunctions}

The introduction of the \DN{Post-entity-attribute} implies adjusting
the functionality of the \textbf{O} function accordingly. The same
modifications can be applied to the formal parameter of actions (not
shown here).


%% file: dynsem.tex
\section{The orchestration logic}
\label{sec:dynsem}
In this section, we present an overview of the reactive computation
model of the orchestration layer. We then formalize it by defining its
semantic algebras, its abstract syntax and its valuation functions.

\subsection{A simple reactive computation model}
Pantagruel is a reactive programming language in that a program constantly
interacts with the outside, reacting to \emph{context changes} that
are observed by changes occurring on the sensed data and the
attributes provided by entities.  A program execution therefore
corresponds to an infinite loop. Each iteration is called an
\emph{orchestration step}, where each step consists of executing
actions on entities according to some conditions on the sensed data or
the attributes of these entities. Each orchestration step produces a
new state reflecting the actions performed.

To facilitate reasoning about the orchestration steps of a program,
we assume :

\begin{enumerate}
\item[i)]\emph{Discrete time and atomic execution.} Discrete time is
  modeled as clock ticks. Each tick corresponds to an orchestration
  step, where all the rules are evaluated within a tick. Invoked
  actions are executed until completion. In doing so, an orchestration
  step is executed with respect to a logical cycle, abstracting over
  time as well as implementation of the actuating capabilities of
  entities.

\item[ii)]\emph{Noninterfering parallelism.} Parallelism assumes that
  all the rules whose conditions are satisfied at a given
  orchestration step, are executed with the same state, making this
  execution simultaneous. Furthermore, we assume a noninterfering
  parallelism, as defined by Schmidt~\cite{davidschmidt} (Chapter~5):
  two rules are executed independently of each other within an
  orchestration step, producing two disjoint, partial states. At the
  end of the execution, states are joined into a final state without
  deleting each other's effects.
  Checking conflicting states that would occur in interfering
  parallelism is left for another paper.

\end{enumerate}


\subsection{Semantic algebras}
\label{sec:dynalgebras}
\paragraph{Interface-centric behavior.} 
The interface abstraction is used to select entities on which an
action needs to be triggered. Such an abstraction also enables
applications to deal with the disappearance or appearance of new
entities while they are executed. In doing so, dynamicity is taken in
account by our language.
The interface-based selection mechanism is ensured by the use of
entity variables involved in rules. For example, Rule (1) of
Figure~\ref{fig:orchestration-layer} declares an \pre{m} entity variable
of \pre{MotionDetector} interface.
To use interface abstraction in rules, we define the ~\DN{Env-entity}
domain as in Figure~\ref{fig:storestruct}, part~\ref{dom:env-entity}:
it maps the identifier of an entity variable to a reference
\DN{Reference} of an entity. The \DN{Reference} domain is a
disjoint union of an \DN{Interface}, reflecting the interface of an
entity that may match a rule, and of an \DN{Instance},
reflecting the entity on which a rule eventually
applies. Entity variables can only be mapped to current entities after
events and actions are evaluated (we examine this point in
Section~\ref{sec:rule-evaluation}).  As a consequence, to evaluate
events in a rule with respect to interface abstraction, we define the
intermediate domain \DO{$\BB$-function} of boolean functions, as
in Figure~\ref{fig:storestruct}, part~\ref{domain:bool}. Specifically,
it needs an environment \DN{Env-entity} to evaluate a boolean
operation. Boolean operators \DO{and$_\rho$} and \DO{or$_\rho$}
are defined accordingly.

Entity variables are inside the scope of a rule; this
makes our language be a block-structured language, as formalized by
Schmidt~\cite{davidschmidt}, Chapter~7: a rule is a block, which scope
is represented by the \DN{Env-entity} domain. Our approach is similar
to the \DN{Location}-based semantics (Chapter~3): a reference has the
same role as a location, holding the denotation of variable
identifiers.
However, in contrast to the \DN{Location} domain, the space of values
covered by entity variables in a rule is constant and restricted to
the elements present in the global store of entities, hence the use
of the specific \DN{Reference} domain.

\paragraph{Context change.}
To represent the reactive nature of Pantagruel, we leverage the
context data provided by the entities of a specification,
focusing on \emph{what} data are available at each orchestration step,
in order to abstract over the way these data are acquired.
The events produced by entities, along with their attributes,
represent the runtime state of a Pantagruel program, as defined by the
\DN{Store} domain (part~\ref{dom:stores} of Figure~\ref{fig:speclayer-semantic-algebras}); orchestration steps are enabled by changes in this
state.  To express changes due to new events, two stores must be
involved, enabling to compare event values between two iterations of
the reactive loop. We make this dual-store strategy explicit with the
compound domain \DO{DualStore}, illustrated in
Figure~\ref{fig:storestruct}, part~\ref{domain:store} .

\begin{figure}[h!tp]
\begin{SemanticAlgebra}
\vspace{-1em}\DomainTitle{Dual stores}\label{domain:store}
\DomainDefNoOp{$\delta$}{DualStore}{Store \mtimes Store}
\DomainTitle{References and environments of entities}
\label{dom:env-entity}
\DomainDef{$f$}{Reference}{Interface + Instance}
\Where{Interface}{Id \xand Instance = Id }
\vspace{.2em}
\DomainDef{$\rho$}{Env-entity}{Id \mto Reference} 
\OPS 
\OperationDefMath{instantiate}{Store \mto Env-entity \mto \PowerSet{Env-entity}}
{\param{\sigma}\param{\rho}\SetOfMath{\param{i}
  ((\rho ~i) ~\DO{equals}~ \IN{Interface}{(\sigma~j)\dasub 1})
  \mto (\IN{Instance}{j}) \xelse (\rho~i)}{j \in Id}}
\DomainTitle{Boolean functions}
\label{domain:bool}%
\DomainDef{$b_\rho$}{$\BB$-function}{Env-entity \mto Tr}
\OPS
\OperationDef{and$_\rho$}{$\BB$-function \mto $\BB$-function \mto Env-entity \mto Tr}
{\param{b_1}\param{b_2}\param{\rho} $(b_1~\rho) \DO{and} (b_2~\rho)$  \NOTE{same for \DO{or$_\rho$}}}\vspace{-.5em}
\end{SemanticAlgebra}
\caption{Semantic algebras for the orchestration layer}
\label{fig:storestruct}
\end{figure}

\subsection{Definition of the orchestration layer}
\label{sec:orchestration-syntax}
The syntax of the orchestration layer is given in
Figure~\ref{fig:astdyn}. An orchestration program P is given as a
specification S and a set K of rules R, relating events W to action
calls C. We examine the language constructs through the valuation
functions, described in the following sections.

\begin{figure}[h!tp]
\vspace{0pt}\begin{minipage}[t]{.3\textwidth}
\begin{syndom}
\DOMAIN{P}{Program}
\DOMAIN{K}{Rule-block} 
\DOMAIN{R}{Rule }
\DOMAIN{W}{Event}
\DOMAIN{C}{Action-call}
\DOMAIN{D}{Declaration}
\DOMAIN{B}{Boolean-expression}
\DOMAIN{F}{Filter}
\DOMAIN{X}{Expression}
\end{syndom}
\end{minipage}\vspace{0pt}
\begin{minipage}[t]{.6\textwidth}
\begin{syndom}
\DOMAIN{I}{Identifier}
\vspace{8.5em} 
\end{syndom}
\end{minipage}
\begin{grammar}
\RULE{P}\CASE{\NT S; \NT K}
\RULE{K}\CASE{rules \NT R end}
\RULE{R}\CASE{\NT{R$_1$} ; \NT{R$_2$} \OR when \NT{W} trigger \NT{C} end}
\RULE{W}
\CASE{\NT{W$_1$} or \NT{W$_2$}
\OR \NT{W$_1$} and \NT{W$_2$} 
\OR event \NT{I$_e$} from \NT D \NT{(}with \NT F\NT{)$^?$} \NT B
}
\RULE{C}
\CASE{\NT{C$_1$} {\small $\|$} \NT{C$_2$}
\OR \NT{C$_1$} , \NT{C$_2$}
\OR action \NT{I$_a$}\,(\NT{X}) on \NT D \NT{(}with \NT F \NT )$^?$} 
\RULE{D}\CASE{\NT{I$_o$} : \NT{I$_c$} \OR \NT{I$_o$}}
\RULE{B}\CASE{value changed \OR value = \NT X}
\RULE{F}\CASE{\NT {I$_p$} = \NT X}
\RULE{X}\CASE{\NT N \OR \NT{I$_o$}. \NT{I$_x$}}
\end{grammar}
\caption{Abstract syntax for the orchestration language}
\label{fig:astdyn}
\end{figure}

\subsubsection{Top-level evaluation : the reactive loop} 
The \textbf{P}  valuation function of a Pantagruel program is depicted
at the top of Figure~\ref{fig:dynsce}.  We represent the reactive loop
as a function from the domain of infinite traces of input stores
($\DN{Store}^\omega$) to its output counterpart ($\DN{Store}^\omega$).
An input store $\sigma^{\pre{i\!n}}$ contains event values pulled from
the outside and captured by the available entities as well as new
entities deployed at runtime.

An output store $\sigma$ results from the evaluation step of a rule
set, as described below.  The initial store $\sigma_{0}$ is defined
from the specification and has no preceding store ($\sigma_{-1}$ is
the empty store). This representation using traces conforms to usual
definitions of reactive languages, such as encountered
in~\cite{BenvenisteGSS92, Brogi97modelingcoordination, Kahn74}.

The evaluation of a program is as follows. The \textbf{S} valuation
function first builds a specification, consisting of an interface
environment $e$ and a store $\sigma_0$ containing the initial values
of the entity attributes. They are given as parameters of the rule
valuation function \textbf{K} along with two store snapshots:
$\sigma_{t-1}$ at time $t-1$, and $\sigma_t'$ at time $t$, including
events values pulled from the outside and new entities. These changes
from the external environment are abstracted by the
\DO{update-external} function.  The evaluation of \textbf{R} results
in a new store, updating the current store $\sigma_t'$ to produce the
store at $t+1$ with the \DO{update-internal} function (its definition
is not shown here). This function is also in charge of setting back to
the undefined value 
all the implicit events that were present in the current store $\sigma_t'$.

%
\begin{figure}[h!tp]
\FunDecl{P}{Program}{Store$^\omega$\mto Store$^\omega$}
\begin{ValuationFunctions}
 	\FunDef{P}{S; K}
    {\param{\Sigma}
        \LetMath{\mtup{e, \sigma_0}}{\FC{S}}
        \In{\SetOfMath[t \in \NN]{
            (\DO{update-internal}~e~(\FC{R}e~\mtup{\sigma_{t-1}, \sigma_{t}'})~\sigma_{t}')}%
            {\sigma^{\pre{i\!n}}_t \in \Sigma, \sigma_t' = (\DO{update-external}~\sigma^{\pre{i\!n}}_t~\sigma_t)}}}
\\\xwhere \sigma_{-1} = \DO{newstore}
\\ \xand \DO{update-external}: \DN{Store} \to \DN{Store} \to \DN{Store}
\\ \xand \DO{update-internal}: \DN{Env-interface} \to \DN{Store} \to \DN{Store} \to \DN{Store}

\end{ValuationFunctions}
\FunDecl{K}{Rule-block}{Env-interface \mto DualStore \mto Store}\\\vspace{-.7em}
\FunDef{K}{\kw{rules} R \kw{end}}{\FC{R}}
\caption{Valuation function of the top-level program}
\label{fig:dynsce}
\end{figure}

\subsubsection{Rule evaluation}
\label{sec:rule-evaluation} The {\bf R} function requires an
environment $\rho$ and a dual store $\delta$ as illustrated in
Figure~\ref{fig:dynrule}. Rule events are evaluated with the {\bf W}
function with respect to the dual-store strategy. Their evaluation
produces an entity environment $\rho_e$, mapping variables, declared
in the event definition, to a reference (\DN{Reference}) to an current
entity. Event evaluation also produces a boolean function whose
execution is done once the $\rho_e$ environment is fully
defined. Actions are then evaluated with the {\bf C} function, taking
$\rho_e$, and updating it into $\rho_a$ with new variables appearing
in actions. Both $\rho_e$ and $\rho_a$ are in the scope of the rule
block~\textbf{when}~...~\textbf{end}.

\paragraph{Rule as a pattern.}
At this stage of the rule evaluation process, variables declared in
the $\rho_a$ environment do not refer to current entities yet,
preventing the action evaluation to complete.  We thus need to
``instantiate'' $\rho_a$ by mapping each of its variables to a current 
entity of the specification.
In fact, more than one entity for each variable of an entity
environment may be concerned by an event (an action, respectively).
As a result, more than one instantiation of environment may match the
events (the actions, respectively).
We thus need to construct a set of such entity environments, mapping
entity variables to identifiers of current entities. This is done by
defining the set $P$: given an entity variable $i_v$ declared with
interface $i_f$, it builds all the environments, such that for each
environment an entity of the same interface $i_f$ is mapped to $i_v$.
Each resulting environment is then given as an argument to the partly
evaluated actions, which return a store.
%
Stores produced by the actions are finally joined with
\emph{join}$^*$.

\begin{figure}[h!tp]
\FunDecl{R}{Rule}{Env-interface \mto DualStore \mto Store}
\begin{ValuationFunctions}
  \FunSeq{R}{\param{e}\param{\delta}%
    (\FCI{R}{1} e~ \delta) \DO{join}
    (\FCI{R}{2} e~ \delta)}\\
  \FunDef{R}{\kw{when} W \kw{trigger} C \kw{end}}
  {\param{e}\param{\mtup{\sigma_{t-1}, \sigma_t}}
      \xlet \mtup{\rho_e, b} = (\FC{W}\mtup{\sigma_{t-1}, \sigma_t}~\DO{empty}~(\param{\rho}\DN{true})) \xin\\\hspace{2em}
      \IndentedFormula{%
        \xlet \mtup{\rho_a, a} = (\FC{C}e~\sigma_t ~\rho_e%
        ~(\param{\rho}\DO{newstore})) \xin\\
        \xlet P = \DO{instantiate} ~\sigma_t~\rho_a \xin\\
       \DO{join$^*$}\SetOfMath{ 
         ((b~\rho_{inst})\to (a~\rho_{inst}) \xelse newstore)}{ \rho_{inst}
\in P }}}
\end{ValuationFunctions}
\caption{Valuation function of a rule}
\label{fig:dynrule}
\end{figure}

\subsubsection{Event evaluation} 
\label{sec:event-evaluation}

The \textbf{W} function evaluates events with a dual store $\delta$
(in ``read-only'' mode), an environment $\rho$, and a boolean function
$b$. The functionality of \textbf{W} shows that an event produces a
pair of functions: an environment and a boolean function. The
environment is updated by the \textbf{D} function, and the boolean
function is defined by the \textbf{B} function.  Specifically, the
\textbf{D} function is interpreted as a variable declaration when it
involves an entity variable, augmenting an environment.
The \textbf{B} function evaluates the B test condition on an event,
which is either of the form \kw{value =} X or \kw{value changed}.
Note that other comparison operators can be envisioned. We omit them
for simplicity.

The evaluation of B depends on an entity environment which is
completed after the evaluation of all the combined W events.  As a
result, \textbf{B} produces an intermediate evaluation of the test
condition, waiting for an environment to complete. This intermediate
evaluation is specified by a \DN{$\BB$-function}, requiring an
environment to produce a boolean value.
The \textbf{B} function also depends on the dual store: when B is of
the form \kw{value =} X, it yields true if the value of event
\FID{I$_e$} equals \FC{X} at time $t$ and not at $t-1$. When B is of
the form \kw{value changed}, it yields true as often as the event
value changes, \ie when its value at $t$ is different from its value at
$t-1$.

Finally, to be concerned by an event, an entity may further need to
satisfy an extra condition.  This condition is evaluated by the
\textbf{F} function, which is similar to \textbf{B}: instead of
testing an event value, it tests an attribute value, and produces a
\DN{$\BB$-function} denoting this test. We ignored the case where the
``\textbf{with} F'' term does not appear, for it is equivalent to
setting ($\FC{F}i~\sigma$) to the constant value (\param{\rho}true).

\begin{figure}[h!tp]
\FunDecl{W}{Event}%
{DualStore \mto Env-entity \mto \DN{$\BB$-function} \mto $($Env-entity \mtimes \DN{$\BB$-function}$)$}
\begin{ValuationFunctions}
\FunDef{W}{W$_1$ \kw{and} W$_2$}
{\param{\delta}\param{\rho}\param{b}\\\xtab
  \IndentedFormula{
    \xlet \mtup{\rho_1, b_1} = (\FCI{W}{1}\delta~ \rho~b) \xin
    (\FCI{W}{2}\delta~\rho_1~b_1)}}
\\
\FunDef{W}{W$_1$ \kw{or} W$_2$}
{\param{\delta}\param{\rho}\param{b}\\\xtab
  \IndentedFormula{
    \xlet \mtup{\rho_1, b_1} = (\FCI{W}{1} \delta~ \rho~b) \xin\\
    \xlet \mtup{\rho_2, b_2} = (\FCI{W}{2} \delta~ \rho_1~b)
    \xin \mtup{\rho_2, (\DO{or$_\rho$}~b_1 ~b_2)}}}
\\
\FunDef{W}{\kw{event} I$_e$ \kw{from} D \kw{with} F B}
{\param{\delta}\param{\rho}\param{b}\\\xtab
  \IndentedFormula{\xlet \mtup{v, \rho'} = \FCOMP{D}{D}\rho \xin\\
  \xlet b_\rho = \param{\rho}\xcases (\DO{access}~v~\rho) \xof \\\xtab\xtab
  \IndentedFormula{\xchoice{\IS{Instance}{i_o}}{
    \IndentedFormula{\xlet x_e = \param{\sigma}(\DO{access-event}\FIDI{I}{e}~i_o~\sigma)
  \xin\\\hspace{.25em}
  (\DO{and$_\rho$}~(\FC{F}i_o~\delta\dasub 1)~(\DO{and$_\rho$}~(\FC{B}~x_e~\delta)~b) ~\rho)}}
\\\xelse\xchoice{\IS{Interface}{i_f}}{\DO{false}}~\xend \hspace{3em}\SEMNOTE{----- ``no entity was found \mto no event is caught''}}
\\ \xin \mtup{\rho', b_\rho}}}\\

\end{ValuationFunctions}

\FunDecl{D}{Declaration}
{Env-entity \mto $($Id \mtimes Env-entity$)$}
\begin{ValuationFunctions}
  \FunDef{D}{I$_v$:I$_f$}{\param{\rho} %
    \mtup{\FIDI{I}{v}, (\DO{update}~\FIDI{I}{v}~\IN{Interface}{\FIDI{I}{f}}~\rho)}}
  \\
  \FunDef{D}{I$_s$}{\param{\rho}\mtup{\FIDI{I}{s},(
\FIDI{I}{s} \in \xdomain(\DO{Store}) 
    \to \DO{update}\FIDI{I}{s}~\IN{Instance}{\FIDI{I}{s}}~\rho \xelse
   \rho}}\\
\end{ValuationFunctions}

\FunDecl{B}{Boolean-expression}{$($Store \mto Value$)$ \mto DualStore \mto $\BB$-function}
\begin{ValuationFunctions}
  \FunDef{B}{\kw{value =} X}{
    \param{x}\param{\delta}
    \xlet b_\rho = \param{\sigma}\param{\rho}(x~ \sigma) \DO{~eq~} (\FC{X}\sigma~\rho) \xin
      \\\xtab\DO{not}(b_\rho~\delta\dasub 1) \DO{~and$_\rho$~} (b_\rho~\delta\dasub 2))}
  \\
  \FunDef{B}{\kw{value changed}}{\param{x}\param{\delta}\param{\rho} %
    ((x~\delta\dasub 1) \DO{neq} (x~\delta\dasub 2))}
\end{ValuationFunctions}

\FunDecl{F}{Filter}{Id \mto Store \mto $\BB$-function}
\begin{ValuationFunctions}
   \FunDef{F}{I$_p$ \kw{=} X}{
     \param{i_o}\param{\sigma}
     \param{\rho}
     (\DO{access-attribute}\FIDI{I}{p}~i_o~\sigma)
     \DO{~eq~}~(\FC{X}\sigma~\rho)}
\end{ValuationFunctions}

\FunDecl{X}{Value}{Store \mto Env-entity \mto Value}
\begin{ValuationFunctions}
\FunDef{X}{N}{\param{\sigma}\param{\rho} \FC{N}}\\
\FunDef{X}{I$_v$.I$_x$}{\param{\sigma}\param{\rho}%
  \xcases (\DO{access}\FIDI{I}{v}~\rho) \xof
  \\\xtab
  \xchoice{\IS{Instance}{i_o}}(\IfThen{\FIDI{I}{x} \in \xdomain(\DN{Entity-event})}{\DO{access-event} \FIDI{I}{x}~i_o~\sigma \xelse \DO{access-attribute} \FIDI{I}{x}~i_o~\sigma})
  \\\xtab\xelse \xchoice{\IS{Interface}{i_f}}{\DN{Undefinedvalue}}~\xend}
\vspace{-1.5em}
\end{ValuationFunctions}
\caption{Valuation functions for events}
\label{fig:dynpred}
\end{figure}

\subsubsection{Action evaluation} 
\label{sec:action-evaluation}
The semantic equations for the \textbf{C} valuation function are
defined in Figure~\ref{fig:dynact}.  \textbf{C} requires an
environment of interfaces, an environment of entities, and a store.
The functionality of \textbf{C} shows that an action produces a pair
of functions: an environment and a function waiting for an environment
to produce a store.
The definition of \textbf{C} mirrors the one of \textbf{W}: similarly
to events, actions can declare variables, which are evaluated by the
\textbf{D} function, enriching the environment given as a parameter
to \textbf{C}.  Similarly to \textbf{W}, \textbf{C} depends on an
environment which is completed after the evaluation of all the
combined C actions; the invoked action identified by \FIDI{I}{a}
produces an intermediate evaluation of this statement, waiting for an
environment argument to complete. This intermediate evaluation is
specified by the intermediate domain \DN{Env-entity \mto Store}
(mirroring the \DN{$\BB$-function} domain for events), requiring an
environment to produce an updated store, reflecting the effect of the
\FIDI{I}{a} action. In our language semantics, we abstract over the
implementation details through the action parameter, by making its
actual value corresponds to the value of the produced
effect. Specifically, it corresponds to the value of the implicit
event generated by the invocation of the action. The store is thus
updated by mapping this actual parameter to an event of the name of
the action, as specified by the \textbf{A} function
(Figure~\ref{fig:speclayer-valuation-functions}).



When actions are executed in parallel, the final store is built by
joining the partial stores produced by these actions (see
Figure~\ref{fig:storestruct} for the\DO{join}operation). In contrast,
a sequential execution builds the argument of the second action by
cumulating the partial store produced by the first action and the
current store.

Note that this process assumes that the scope of a sequence of actions
is bound to the environment in which the concerned actions apply.  In
other words, the order of actions defined sequentially over an entity
variable is preserved only for the actions of the current entity 
to which the variable is mapped.

\begin{figure}[htp!]
  {\small \FunDecl{C}{\small Action}%
  {\! Env-interface \mto Store\! \mto Env-entity \mto $\!($Env-entity \mto Store$)$ 
      \!\mto\! $($Env-entity \!\mtimes\!\! $($Env-entity\! \mto Store$)\!)$}}
  \begin{ValuationFunctions} 
\FunDef{C}{C$_1$ $\|$ C$_2$}{\param{e}\param{\sigma}\param{\rho}\param{f_\sigma}%
  \\\xtab\IndentedFormula{
    \xlet \mtup{\rho_1, 
      f_{\sigma_1}} = (\FCI{C}{1}e~\sigma~\rho~f_\sigma) \xin\\
    \xlet \mtup{\rho_2, 
      f_{\sigma_2}} = (\FCI{C}{2}e~\sigma~\rho_1~f_\sigma) \xin %
    \mtup{\rho_2, 
      \param{\rho}(f_{\sigma_2}~\rho) \DO{join} (f_{\sigma_1}~\rho)}}}
\\
\FunDef{C}{C$_1$ \kw , C$_2$}{\param{e}\param{\sigma}\param{\rho}\param{f_\sigma}%
  \\\xtab\IndentedFormula{
    \xlet \mtup{\rho_1, f_{\sigma_1}} = (\FCI{C}{1}e~\sigma~\rho~f_\sigma) 
    \xin (\FCI{C}{2}e~\sigma~\rho_1~f_{\sigma_1})}}
\\
\FunDef{C}{\kw{action} I$_a$ \kw ( X \kw ) \kw{on} D \kw{with} F}
{ \param{e}\param{\sigma} \param{\rho} \param{f_\sigma}
   \\\xtab\xlet \mtup{v, \rho'} = (\FCOMP{D}{D}\rho) \xin
    \\\xtab\xlet \sigma_\rho = \param{\rho}\xcases (\DO{access}~v~\rho) \xof\\
    \xtab\xtab \IndentedFormula{\xchoice{\IS{Instance}{i_o}}{
    \IndentedFormula{\xlet a_\rho = 
      (\DO{access-action}\FIDI{I}{a}~i_o~e~\sigma)
      \xin\\
      \xlet f_\sigma' = \param{\rho}(a_\rho~i_o~(\FC{X}\rho~\sigma))~(f_\sigma~\rho) \xin\\
      \IfThen{\FC{F}~i_o~\sigma~\rho}{(f_\sigma'~\rho)\xelse
        (f_\sigma~\rho)})}}
  \\\xelse\xchoice{\IS{Interface}{i_f}}{f_\sigma~\rho}~ \xend  \hspace{3em}\SEMNOTE{----- ``no entity was found \mto no store update''}}
  \\\xtab \xin \mtup{\rho', \sigma_\rho}}\vspace{-1.5em}
\end{ValuationFunctions}
\caption{Valuation functions of actions}
\label{fig:dynact}
\end{figure}

\input{semexample}


%% file: semexample.tex
\subsection{Applying the semantics to an example}
The static semantics analysis of Rule (1) in the
program example  of Figure~\ref{fig:orchestration-layer} is presented
in Figures~\ref{fig:example-semantics-one} and \ref{fig:example-semantics}.
This example demonstrates that the intended behavior of a Pantagruel
program is covered by its semantics. In particular, as illustrated in
Figure~\ref{fig:example-semantics}, the store $\sigma_{R1}$ produced
by the execution of Rule (1), contains two events, reflecting the
execution of the \texttt{switch} action on the \texttt{l10} and
\texttt{l11} lights, located in room \texttt{101}, where the
\texttt{m10} motion detector detected an event from $\sigma_1$ to
$\sigma_2$. Most \xcases constructs for handling \DN{Reference}
elements are skipped in our example, in order to keep the
clarity of the illustration.


\newcommand{\tikzmark}[1]{\tikz[overlay,remember picture] \node (#1) {};}
\tikzset{arrow/.style={to path={ -- (\tikztotarget)},
decoration={markings,mark=at position 1 with {\arrow[scale=1.5,#1]{>}}},
    postaction={decorate},
    shorten >=0.2pt}}
\newcommand{\DrawArrow}[3][]{%
  \tikz[overlay,remember picture] {\draw[->,arrow, thick, #1] 
    ($(.1,.1ex)+(#2.south)$) to ($(#3.north)+(.19,1.4ex)$);} 
}
\begin{figure}[htp!]
Let \parbox[t]{.9\textwidth}{%
\begin{tabular}[t]{ll}
D$_0$ &=~ \pre{m:MotionDetector}\\
D$_1$ &=~ \pre{l:Light}\\
B$_0$ &=~ \kw{value =} \pre{true}\\
F$_0$ &=~ \pre{room = m.room}\\
W$_0$ &=~ \kw{event} \pre{detected} \kw{from} D$_0$ B$_0$\\
C$_0$ &=~ \kw{action} \pre{switch(true)} \kw{on} D$_1$ \kw{with} F$_0$\\
\end{tabular}
}
\\and an initial store $\sigma_0 = \FIDp{l10} \mapsto \mtup{\FIDp{Light}, \mtup{[\FIDp{room}\mapsto 101]\DO{empty}, \DO{empty}} ...}$ containing the entities of Figure~\ref{fig:specification-layer}.
Moreover, in this example, we assume that a motion was detected from $\sigma_1$ to $\sigma_2$.

~\vspace{.6em}\\
\FCOMP{R}{\kw{when} W$_0$ \kw{trigger} C$_0$}$e ~ \mtup{\sigma_{1}, \sigma_{2}}$\\
=\IndentedFormula{
  \xlet \mtup{\rho_e, b} = 
  (\FCI{W}{0})\mtup{\sigma_1, \sigma_2}~\DO{empty}~(\param{\rho}\;true) \xin\\
  \xlet \mtup{\rho_a, a} = (\FCI{C}{0}e~\sigma_2 ~\rho_e ~(\param{\rho}\DO{newstore}))
  \xin \\
  \xlet P = \DO{instantiate} ~\sigma_2~\rho_a 
  \xin \DO{join$^*$}\SetOfMath{((b~\rho_{inst})\to (a~\rho_{inst}) 
                                     \xelse newstore)}{\rho_{inst} \in P}}
 \vspace{1em}\\
  \noindent Where : \vspace{.5em}\\
  (1)~~\IndentedFormula{
    \FCI{W}{0}\mtup{\sigma_{1}, \sigma_{2}}\DO{empty}~(\param{\rho}\DO{true})\\
    = \FCOMP{W}{\kw{event} \pre{detected} \kw{from} D$_0$ B$_0$}
    \mtup{\sigma_{1}, \sigma_{2}}\DO{empty}~(\param{\rho}\;true)%
    \\
    =\IndentedFormula{
      \tikzmark{LET1}%
      \xlet \mtup{v, \rho'} = \IndentedFormula{
        \FCOMP{\tikzmark{D}D}{\pre{m:MotionDetector}}\DO{empty} \xin\\~\\
        \tikzmark{vD}\mtup{\FIDp{m},
          \underbrace{
            (\DO{update}\FIDp{m}~(\IN{Interface}{\FIDp{MotionDetector}})~\DO{empty})}_{\rho' \textnormal{ equals } \callupdate{\FIDp{m}}{\IN{Interface}{\FIDp{MotionDetector}}}{\DO{empty}}}} \xin
        \\ 
        \xlet b_\rho = \param{\rho}~... \left[\begin{array}{@{}l}
            \xlet x_e = \param{\sigma}(
            \underbrace{
              \DO{access-event}\FIDp{detected}~(\DO{access}~\FIDp{m}~\rho)~\sigma}_{\xlet 
              \textnormal{this be } \xstr{m.detected}\sigma})
            \xin\\~\\
            \tikzmark{vLET1} 
              \DO{and$_\rho$}~(\tikzmark{B}\FCI{B}{0}~x_e~\mtup{\sigma_{1}, 
                \sigma_{2}})~(\param{\rho}\DO{true})~\rho~^{\DO{(note)}}
          \end{array}  
        \right. }\\~\\
      \hspace{16.5em}\IndentedFormula{
        \tikzmark{vB}\FCOMP{B}{\kw{value =} true}\\
        \begin{array}{@{}l@{}}=\\~\\~\\\end{array}\tikzmark{vB1}\left[\begin{array}{@{}l} \xlet b_\rho = 
            \param{\sigma}\param{\rho}
            \underbrace{(x_e~ \sigma)}_{\xstr{m.detected}\sigma}
            ~\DO{eq}~
            \underbrace{(\FCOMP{X}{true}\sigma~\rho)}_{true}
            \xin
            \\\DO{not}(b_\rho~\sigma_1) \DO{~and$_\rho$~} (b_\rho~\sigma_2)
          \end{array}\right. 
        \\
        = \DO{not}(\xstr{m.detected}\sigma_1)~\DO{and$_\rho$}~(\xstr{m.detected}\sigma_2)
      }
      \vspace{1em}\\ = \mtup{\underbrace{\callupdate{\FIDp{m}}{\IN{Interface}{\pre{MotionDetector}}}{\DO{empty}}}_{\rho_e}, 
        \underbrace{\DO{not}(\xstr{m.detected}\sigma_1)~\DO{and$_\rho$}~(\xstr{m.detected}\sigma_2)}_{b}}}}
  \DrawArrow{D}{vD}
  \DrawArrow{B}{vB}
  ~\\\xnote{(note) no F filter in W$_0$ is 
    equivalent to the ``\kw{with} true'' filter, so we skipped it here}
  \vspace{-2em}
\caption{Semantics applied on Figure~\ref{fig:orchestration-layer}}
\label{fig:example-semantics-one}
\end{figure}

\begin{figure}[h!tp]
  (2)~~\IndentedFormula{
    \FCI{C}{0}e~\sigma_2~~\rho_e~(\param{\rho'}\DO{newstore})\\
    =\FCOMP{C}{\kw{action} \pre{switch(true)} \kw{on} D$_1$ \kw{with} F$_0$}\\
    =  \xlet \mtup{v, \rho'} = \tikzmark{D1}\FCOMP{D}{\pre{l:Light}}\rho_e \xin 
    \\~\\\hspace{6.5em}\tikzmark{vD1}%
    \mtup{\FID{l}, \underbrace{\DO{update}\FIDp{l}~(\IN{Interface}{\FIDp{Light})}~\rho_{e}}
      _{\callupdate{\FIDp{l}}{\IN{Interface}{\FIDp{Light}}}{\rho_e}}} \xin 
    \\
    \hspace{4em}\xlet \sigma_\rho = \param{\rho}~ ... \IndentedFormula{
      \left[ 
        \begin{array}{@{}l}
    \xlet a_\rho = 
    \underbrace{(\DO{access-action}\FIDp{switch}~(access~\FID{l}~\rho)~e~\sigma)}_{
(\param{i_o}\param{x}(\DO{update-event}\FIDp{switch}~i_o~x))(\DO{access}~\FIDp{l}~\rho)
}) \xin\\
          \xlet f_\sigma' = \param{\rho}
          \underbrace{(a_\rho~(\DO{access}~\FIDp{l}~\rho)~(\FC{X}\rho~\sigma))~(f_\sigma~\rho)}
\vspace{-.4em}\\{\hspace{5em} }_{[(\DO{access}~\FIDp{l}~\rho) \mapsto \mtup{..., \callupdate{\FID{switch}}{true}{empty}}] \DO{newstore}, \textnormal{and let this be  $\sigma_2'$}}
\\
\IfThen{~\tikzmark{F}\FCI{F}{0}~\FID{l}~\sigma_2~\rho}{(f_\sigma'~\rho)\xelse \DO{newstore}}
            
          \end{array}
      \right.
      \\~\\\hspace{2em}\IndentedFormula{\tikzmark{vF}\FCOMP{F}{\pre{room} \kw{=} \pre{m.room}}(access~\FID{l}~\rho)~\sigma_2~\rho
        ~\\
        \\ \tikzmark{vF1}%
        \underbrace{\DO{access-attribute}\FIDp{room}~(\DO{access}~\FID{l}~\rho)~\sigma_2}_{\textnormal{let this be} \xstr{l.room}\rho~\sigma_2}~\DO{eq}~
        \IndentedFormula{\tikzmark{X}\FCOMP{X}{\pre{m.room}}\rho~\sigma_2\vspace{1em}\\
          \tikzmark{vX}(3) \textnormal{-- see below}}
      }}
    \vspace{1em}\\=  \mtup{
      \underbrace{\callupdate{\FIDp{l}}{\IN{Interface}{\FIDp{Light}}}{\rho_e}}_{\rho_a}, 
      \underbrace{\param{\rho}((\xstr{l.room}\rho~\sigma_2) \DO{~eq~} \xdots{(3)}) \to \sigma_2' \xelse \DO{newstore})}_{a}
    }}

  \vspace{1em}\xtab \textnormal{where : }\vspace{1em}\\
    (3)~~\IndentedFormula{\FCOMP{X}{\pre{m.room}}\rho~\sigma_2\\
    = \xcases \DO{access}\FIDp{m}~\rho \xof
    \xchoice{\IN{Instance}{i_o}}{(\DO{access-attribute}\FIDp{room}~i_o~\sigma_2)}
    \\\hspace{9.7em}\xelse \xchoice{\IN{Interface}{\FIDp{MotionDetector}}}{\DN{Undefinedvalue}}
  }
  \DrawArrow{D1}{vD1}
  \DrawArrow{F}{vF}
  \DrawArrow{vF}{vF1}
  \DrawArrow{X}{vX}

\noindent Where : \vspace{.2em}\\
  (4)~~\IndentedFormula{\DO{instantiate} ~\sigma_2~\rho_a \\
  = \xlet P = \{ 
      \begin{array}[t]{@{}l@{}}
        \FIDp{m} \mapsto \IN{Instance}{j}, 
        {\small \textnormal{ with } j \in \{\pre{m10, m20}\},
          \textnormal{ and } (\sigma_2~j)\dasub 1 = \FIDp{MotionDetector}}\\
        \FIDp{l} \mapsto \IN{Instance}{k}, 
        {\small \textnormal{ with } k \in \{\pre{l10, l11, l20}\},
          \textnormal{ and } (\sigma_2~k)\dasub 1 = \FIDp{Light}}\} \xin \\
        \end{array}  
    \\\DO{join$^*$}\SetOfMath{((b~\rho_{inst})\to (a~\rho_{inst}) \xelse newstore)}{ \rho_{inst}
\in P}, 
    }
    \vspace{1em}\\
    Now, the $b$ and $a$ functions built in (1) and (2) can be fully evaluated with
    \eg : 
    \\i) \IndentedFormula{P = \{\rho_1, \rho_2, ... \}, \textnormal{ such that }\\\hspace{1em}
      \rho_1 = \{\FIDp{m} \mapsto \IN{Instance}{\pre{m10}},
             \FIDp{l} \mapsto \IN{Instance}{\pre{l10}}\},\\\hspace{1em} 
             \rho_2 = \{\FIDp{m} \mapsto \IN{Instance}{\pre{m10}},
             \FIDp{l} \mapsto \IN{Instance}{\pre{l11}}\},\emph{etc.}}
    \vspace{1em}\\
    ii) \IndentedFormula{
      \sigma_1 = \IndentedFormula{
        \left[\FIDp{m10} \mapsto [\FIDp{room} \mapsto \FIDp{101}][\FIDp{detected} \mapsto \texttt{false}]\DO{empty}\right]
        \\ \left[\FIDp{l10} \mapsto [\FIDp{room} \mapsto \FIDp{101}]\DO{empty}\right]\DO{newstore}
        \\ \left[\FIDp{l11} \mapsto [\FIDp{room} \mapsto \FIDp{101}]\DO{empty}\right]\DO{newstore}} 
      \\\sigma_2 = [\FIDp{m10} \mapsto [\FIDp{room} \mapsto \FIDp{101}][\FIDp{detected} \mapsto \texttt{\color{red}{true}}]\DO{empty}]~
             ... \textnormal{same as } \sigma_1... ]  \DO{newstore}
    }
    
    ~\vspace{.6em}\\As a result, the store $\sigma_{R1}$, resulting from the application of \DN{join}$^*$ equals :
    \\$\sigma_{R1} = [$\IndentedFormula{
      \FIDp{l10} \mapsto \mtup{\FIDp{Light},\xdots{\DO{entity-attributes}},[\FIDp{switch} \mapsto true]\DO{empty}}][\\
\FIDp{l11} \mapsto \mtup{\FIDp{Light},\xdots{\DO{entity-attributes}},[\FIDp{switch} \mapsto true]\DO{empty}} ]\DO{newstore}
    }
  \caption{Semantics applied on Figure~\ref{fig:orchestration-layer} (continued (2))}
  \label{fig:example-semantics}
\end{figure}

%% file: conclu.tex
\clearpage
\section{Conclusion and Future Work}
\label{sec:conclu}
In this paper, we presented the realisation of a user-oriented
domain-specific language, named Pantagruel, aimed to ease the
development of entity orchestration applications. The design of
Pantagruel is supported by a thorough analysis of the entity
orchestration domain, bringing out the domain requirements and the
needs of users. In addition, we used the formal methodology for
language development proposed by David Schmidt~\cite{davidschmidt}. It
identifies the key concepts in language design and
semantics. Throughout the realisation of Pantagruel, we shown how the
denotational semantics of our language reflects the domain analysis,
expressing the user-oriented concepts of the domain as first class
domains of the semantics. Finally, we hope the example of Pantagruel
serves as a basis to help and motivate language developers to provide
a proper semantic definition of their DSL.

Once the semantics of our language is formally defined, it can be used
both to derive implementations (\ie interpreters, compilers) and to
reason about programs by expressing various program analyses. An
interpreter for Pantagruel has been developed in OCaml; it is a
straightforward mapping of the formal definition. An implementation of
a compiler towards a Java platform has also been developed based on the
semantics, and was used for a demonstration of Pantagruel at a conference~\cite{dreypercom}.

As for future work, we will use the semantics of
Pantagruel to perform static program analysis, such as conflict
detection when rule execution affects the state of the same
entities. To do so, we can leverage Nakata's work on the While
reactive language~\cite{nakatadsl11}. Nakata uses the Coq proof
assistant to perform property verifications on the While language,
based on a denotational-style definition. We are also studying an
approach to allowing end users to express area-specific properties,
describing the intended behavior of their orchestration application,
in a user-friendly paradigm. Because they are most qualified to
determine what area-specific properties should be satisfied by
orchestration applications,
it is essential to provide them with high-level abstractions to make
accessible the description of area-specific properties. To support the
verification of orchestration applications against these properties,
this work will build upon the formal semantics of Pantagruel. Another
interesting direction based on the denotational semantics of
Pantagruel, is to develop tools such as debuggers and profilers, and
to make them accessible to end users.

\vspace{-.25em}
\subsection*{Acknowledgment}
\vspace{-.4em}
The formal definition of a programming language can be intricate and dense. David Schmidt has made landmark contributions on this topic, providing researchers with a practical methodology to formalize a programming language definition without sacrificing underlying mathematical foundations. His approach makes the design of a language definition systematic, rigorous and tasteful. Resulting definitions are easier to understand by implementers, to use for formal reasoning, and to leverage to introduce non-standard semantics. This paper is a testimony to the fact that the appealing nature of David Schmidt's methodology is strong across generations of researchers in programming languages.




%% file: festschrift2013.bbl
\begin{thebibliography}{10}
\providecommand{\bibitemdeclare}[2]{}
\providecommand{\surnamestart}{}
\providecommand{\surnameend}{}
\providecommand{\urlprefix}{Available at }
\providecommand{\url}[1]{\texttt{#1}}
\providecommand{\href}[2]{\texttt{#2}}
\providecommand{\urlalt}[2]{\href{#1}{#2}}
\providecommand{\doi}[1]{doi:\urlalt{http://dx.doi.org/#1}{#1}}
\providecommand{\bibinfo}[2]{#2}

\bibitemdeclare{article}{BenvenisteGSS92}
\bibitem{BenvenisteGSS92}
\bibinfo{author}{Albert \surnamestart Benveniste\surnameend},
  \bibinfo{author}{Paul~Le \surnamestart Guernic\surnameend},
  \bibinfo{author}{Yves \surnamestart Sorel\surnameend} \&
  \bibinfo{author}{Michel \surnamestart Sorine\surnameend}
  (\bibinfo{year}{1992}): \emph{\bibinfo{title}{A Denotational Theory of
  Synchronous Reactive Systems}}.
\newblock {\sl \bibinfo{journal}{Information and Computation}}
  \bibinfo{volume}{99}(\bibinfo{number}{2}), pp. \bibinfo{pages}{192--230},
  \doi{10.1016/0890-5401(92)90030-J}.

\bibitemdeclare{inproceedings}{Brogi97modelingcoordination}
\bibitem{Brogi97modelingcoordination}
\bibinfo{author}{Antonio \surnamestart Brogi\surnameend} \&
  \bibinfo{author}{Jean{-}Marie \surnamestart Jacquet\surnameend}
  (\bibinfo{year}{1997}): \emph{\bibinfo{title}{Modeling Coordination via
  Asynchronous Communication}}.
\newblock In: {\sl \bibinfo{booktitle}{Coordination '97, LNCS}},
  \bibinfo{publisher}{Springer-Verlag}, pp. \bibinfo{pages}{238--255},
  \doi{10.1007/3-540-63383-9\_84}.

\bibitemdeclare{article}{diaspectse}
\bibitem{diaspectse}
\bibinfo{author}{Damien \surnamestart Cassou\surnameend},
  \bibinfo{author}{Julien \surnamestart Bruneau\surnameend},
  \bibinfo{author}{Charles \surnamestart Consel\surnameend} \&
  \bibinfo{author}{Emilie \surnamestart Balland\surnameend}
  (\bibinfo{year}{2012}): \emph{\bibinfo{title}{{Towards a Tool-based
  Development Methodology for Pervasive Computing Applications}}}.
\newblock {\sl \bibinfo{journal}{{IEEE Transactions on Software Engineering}}}
  \bibinfo{volume}{38}(\bibinfo{number}{6}), pp. \bibinfo{pages}{1445--1463},
  \doi{10.1109/TSE.2011.107}.

\bibitemdeclare{inbook}{conseldsl}
\bibitem{conseldsl}
\bibinfo{author}{Charles \surnamestart Consel\surnameend}
  (\bibinfo{year}{2004}): \emph{\bibinfo{title}{Domain-Specific Program
  Generation; International Seminar, Dagstuhl Castle}}, chapter
  \bibinfo{chapter}{From A Program Family To A Domain-Specific Language}, pp.
  \bibinfo{pages}{19--29}.
\newblock {\sl \bibinfo{series}{Lecture Notes in Computer Science}}
  \bibinfo{volume}{3016}, \bibinfo{publisher}{Springer-Verlag},
  \bibinfo{address}{London, UK}, \doi{10.1007/978-3-540-25935-0\_2}.

\bibitemdeclare{inproceedings}{conselmarlet}
\bibitem{conselmarlet}
\bibinfo{author}{Charles \surnamestart Consel\surnameend} \&
  \bibinfo{author}{Renaud \surnamestart Marlet\surnameend}
  (\bibinfo{year}{1998}): \emph{\bibinfo{title}{Architecturing software using a
  methodology for language development}}.
\newblock In: {\sl \bibinfo{booktitle}{Proceedings of the 10 th International
  Symposium on Programming Language Implementation and Logic Programming,
  number 1490 in Lecture Notes in Computer Science}}, pp.
  \bibinfo{pages}{170--194}, \doi{10.1007/BFb0056614}.

\bibitemdeclare{article}{little-languages}
\bibitem{little-languages}
\bibinfo{author}{Arie \surnamestart van Deursen\surnameend} \&
  \bibinfo{author}{Paul. \surnamestart Klint\surnameend}
  (\bibinfo{year}{1998}): \emph{\bibinfo{title}{Little Languages: Little
  Maintenance}}.
\newblock {\sl \bibinfo{journal}{Journal of Software Maintenance}}
  \bibinfo{volume}{10}(\bibinfo{number}{2}), pp. \bibinfo{pages}{75--92},
  \doi{10.1002/(SICI)1096-908X(199803/04)10:2<75::AID-SMR168>3.0.CO;2-5}.

\bibitemdeclare{article}{annotated}
\bibitem{annotated}
\bibinfo{author}{Arie \surnamestart van Deursen\surnameend},
  \bibinfo{author}{Paul \surnamestart Klint\surnameend} \&
  \bibinfo{author}{Joost \surnamestart Visser\surnameend}
  (\bibinfo{year}{2000}): \emph{\bibinfo{title}{Domain-specific Languages: an
  annotated bibliography}}.
\newblock {\sl \bibinfo{journal}{SIGPLAN Notices}}
  \bibinfo{volume}{35}(\bibinfo{number}{6}), pp. \bibinfo{pages}{26--36},
  \doi{10.1145/352029.352035}.

\bibitemdeclare{techreport}{phddrey}
\bibitem{phddrey}
\bibinfo{author}{Zo{\'e} \surnamestart Drey\surnameend} (\bibinfo{year}{2010}):
  \emph{\bibinfo{title}{{Vers une m{\'e}thodologie d{\'e}di{\'e}e {\`a}
  l'orchestration d'entit{\'e}s communicantes}}}.
\newblock \bibinfo{type}{Ph.D. thesis}, \bibinfo{institution}{PHOENIX - INRIA
  Bordeaux - Sud-Ouest}.

\bibitemdeclare{inproceedings}{dreypercom}
\bibitem{dreypercom}
\bibinfo{author}{Zo{\'e} \surnamestart Drey\surnameend} \&
  \bibinfo{author}{Charles \surnamestart Consel\surnameend}
  (\bibinfo{year}{2010}): \emph{\bibinfo{title}{A visual, open-ended approach
  to prototyping ubiquitous computing applications}}.
\newblock In: {\sl \bibinfo{booktitle}{PerCom Workshops}}, pp.
  \bibinfo{pages}{817--819}, \doi{10.1109/PERCOMW.2010.5470549}.

\bibitemdeclare{article}{jvlc12}
\bibitem{jvlc12}
\bibinfo{author}{Zo{\'e} \surnamestart Drey\surnameend} \&
  \bibinfo{author}{Charles \surnamestart Consel\surnameend}
  (\bibinfo{year}{2012}): \emph{\bibinfo{title}{Taxonomy-driven prototyping of
  home automation applications: A novice-programmer visual language and its
  evaluation}}.
\newblock {\sl \bibinfo{journal}{Journal of Visual Langages and Computing}}
  \bibinfo{volume}{23}(\bibinfo{number}{6}), pp. \bibinfo{pages}{311--326},
  \doi{10.1016/j.jvlc.2012.07.002}.

\bibitemdeclare{inproceedings}{dsl09}
\bibitem{dsl09}
\bibinfo{author}{Zo{\'e} \surnamestart Drey\surnameend},
  \bibinfo{author}{Julien \surnamestart Mercadal\surnameend} \&
  \bibinfo{author}{Charles \surnamestart Consel\surnameend}
  (\bibinfo{year}{2009}): \emph{\bibinfo{title}{A Taxonomy-Driven Approach to
  Visually Prototyping Pervasive Computing Applications}}.
\newblock In: {\sl \bibinfo{booktitle}{DSL '09: Proceedings of the IFIP TC 2
  Working Conference on Domain-Specific Languages}},
  \bibinfo{publisher}{Springer-Verlag}, \bibinfo{address}{Berlin, Heidelberg},
  pp. \bibinfo{pages}{78--99}, \doi{10.1007/978-3-642-03034-5\_5}.

\bibitemdeclare{book}{martinfowler}
\bibitem{martinfowler}
\bibinfo{author}{Martin \surnamestart Fowler\surnameend}
  (\bibinfo{year}{2010}): \emph{\bibinfo{title}{Domain Specific Languages}},
  \bibinfo{edition}{1st} edition.
\newblock \bibinfo{publisher}{Addison-Wesley Professional}.

\bibitemdeclare{article}{dsel}
\bibitem{dsel}
\bibinfo{author}{Paul \surnamestart Hudak\surnameend} (\bibinfo{year}{1996}):
  \emph{\bibinfo{title}{Building Domain-Specific Embedded Languages}}.
\newblock {\sl \bibinfo{journal}{ACM Computing Surveys}} \bibinfo{volume}{28},
  \doi{10.1145/242224.242477}.

\bibitemdeclare{inproceedings}{Kahn74}
\bibitem{Kahn74}
\bibinfo{author}{Gilles \surnamestart Kahn\surnameend} (\bibinfo{year}{1974}):
  \emph{\bibinfo{title}{The Semantics of Simple Language for Parallel
  Programming}}.
\newblock In: {\sl \bibinfo{booktitle}{IFIP Congress}}, pp.
  \bibinfo{pages}{471--475}.

\bibitemdeclare{article}{endusersoftware}
\bibitem{endusersoftware}
\bibinfo{author}{Andrew~J. \surnamestart Ko\surnameend}, \bibinfo{author}{Robin
  \surnamestart Abraham\surnameend}, \bibinfo{author}{Laura \surnamestart
  Beckwith\surnameend}, \bibinfo{author}{Alan \surnamestart
  Blackwell\surnameend}, \bibinfo{author}{Margaret \surnamestart
  Burnett\surnameend}, \bibinfo{author}{Martin \surnamestart Erwig\surnameend},
  \bibinfo{author}{Chris \surnamestart Scaffidi\surnameend},
  \bibinfo{author}{Joseph \surnamestart Lawrance\surnameend},
  \bibinfo{author}{Henry \surnamestart Lieberman\surnameend},
  \bibinfo{author}{Brad \surnamestart Myers\surnameend},
  \bibinfo{author}{Mary~Beth \surnamestart Rosson\surnameend},
  \bibinfo{author}{Gregg \surnamestart Rothermel\surnameend},
  \bibinfo{author}{Mary \surnamestart Shaw\surnameend} \&
  \bibinfo{author}{Susan \surnamestart Wiedenbeck\surnameend}
  (\bibinfo{year}{2011}): \emph{\bibinfo{title}{The state of the art in
  end-user software engineering}}.
\newblock {\sl \bibinfo{journal}{ACM Computing Surveys}}
  \bibinfo{volume}{43}(\bibinfo{number}{3}), pp. \bibinfo{pages}{21:1--21:44},
  \doi{10.1145/1922649.1922658}.

\bibitemdeclare{inproceedings}{liang96moddensem}
\bibitem{liang96moddensem}
\bibinfo{author}{Sheng \surnamestart Liang\surnameend} \& \bibinfo{author}{Paul
  \surnamestart Hudak\surnameend} (\bibinfo{year}{1996}):
  \emph{\bibinfo{title}{Modular Denotational Semantics for Compiler
  Construction}}.
\newblock In: {\sl \bibinfo{booktitle}{European Symposium on Programming}},
  \bibinfo{publisher}{Springer-Verlag}, pp. \bibinfo{pages}{219--234},
  \doi{10.1007/3-540-61055-3\_39}.

\bibitemdeclare{article}{whenandhow}
\bibitem{whenandhow}
\bibinfo{author}{Marjan \surnamestart Mernik\surnameend}, \bibinfo{author}{Jan
  \surnamestart Heering\surnameend} \& \bibinfo{author}{Anthony~M.
  \surnamestart Sloane\surnameend} (\bibinfo{year}{2005}):
  \emph{\bibinfo{title}{When and how to develop domain-specific languages}}.
\newblock {\sl \bibinfo{journal}{ACM Computing Surveys}}
  \bibinfo{volume}{37}(\bibinfo{number}{4}), pp. \bibinfo{pages}{316--344},
  \doi{10.1145/1118890.1118892}.

\bibitemdeclare{inproceedings}{nakatadsl11}
\bibitem{nakatadsl11}
\bibinfo{author}{Keiko \surnamestart Nakata\surnameend} (\bibinfo{year}{2011}):
  \emph{\bibinfo{title}{Resumption-based big-step and small-step interpreters
  for While with interactive I/O}}.
\newblock In: {\sl \bibinfo{booktitle}{DSL '11: Proceedings of the IFIP TC 2
  Working Conference on Domain-Specific Languages}}, pp.
  \bibinfo{pages}{226--235}, \doi{10.4204/EPTCS.66.12}.

\bibitemdeclare{misc}{schmidtintro}
\bibitem{schmidtintro}
\bibinfo{author}{David~A. \surnamestart Schmidt\surnameend}:
  \emph{\bibinfo{title}{An introduction to Programming-Language Semantics}}.
\newblock
  \bibinfo{howpublished}{\url{http://people.cis.ksu.edu/~schmidt/705s13/Lectur%
es/chapter.pdf}}.
\newblock \bibinfo{note}{A revision of the article in the {CRC/ACM} {C}omputer
  {S}cience {H}andbook, 2d ed., 2004.}

\bibitemdeclare{book}{davidschmidt}
\bibitem{davidschmidt}
\bibinfo{author}{David~A. \surnamestart Schmidt\surnameend}
  (\bibinfo{year}{1986}): \emph{\bibinfo{title}{Denotational semantics: a
  methodology for language development}}.
\newblock \bibinfo{publisher}{Allyn and Bacon}, \bibinfo{address}{Boston,
  London}.

\bibitemdeclare{article}{notablepatterns}
\bibitem{notablepatterns}
\bibinfo{author}{Diomidis \surnamestart Spinellis\surnameend}
  (\bibinfo{year}{2001}): \emph{\bibinfo{title}{Notable design patterns for
  domain-specific languages}}.
\newblock {\sl \bibinfo{journal}{Journal of Systems and Software}}
  \bibinfo{volume}{56}(\bibinfo{number}{1}), pp. \bibinfo{pages}{91--99},
  \doi{10.1016/S0164-1212(00)00089-3}.

\bibitemdeclare{article}{tratt08}
\bibitem{tratt08}
\bibinfo{author}{Laurence \surnamestart Tratt\surnameend}
  (\bibinfo{year}{2008}): \emph{\bibinfo{title}{Domain specific language
  implementation via compile-time meta-programming}}.
\newblock {\sl \bibinfo{journal}{ACM Transactions on Programming Languages and
  Systems}} \bibinfo{volume}{30}(\bibinfo{number}{6}), pp.
  \bibinfo{pages}{31:1--31:40}, \doi{10.1145/1391956.1391958}.

\bibitemdeclare{inproceedings}{wangdsl}
\bibitem{wangdsl}
\bibinfo{author}{Qian \surnamestart Wang\surnameend} \& \bibinfo{author}{Gopal
  \surnamestart Gupta\surnameend} (\bibinfo{year}{2005}):
  \emph{\bibinfo{title}{Rapidly prototyping implementation infrastructure of
  domain specific languages: a semantics-based approach}}.
\newblock In: {\sl \bibinfo{booktitle}{SAC}}, pp. \bibinfo{pages}{1419--1426}.

\end{thebibliography}
